\documentclass[%
 aip,
 amsmath,amssymb,
 reprint,%
nofootinbib 
]{revtex4-1}

\usepackage{graphicx}
\usepackage{dcolumn}
\usepackage{bm}

\usepackage[utf8]{inputenc}
\usepackage[T1]{fontenc}
\usepackage{mathptmx}
\usepackage{etoolbox}
\usepackage{braket}
\usepackage{color}
\usepackage{subfigure}
\usepackage{comment}

\newcommand{\wig}[3]{D^{#1}_{#2}(#3)}

\newcommand{\virg}[1]{``#1''}
\newcommand*{\h}{\mathcal{H}}
\newcommand*{\Tr}{\mathrm{Tr}}
\newcommand*{\diff}{\text{d}}
\newcommand*{\coloneqq}{\mathrel{\vcenter{\baselineskip0.5ex \lineskiplimit0pt \hbox{\scriptsize.}\hbox{\scriptsize.}}} =}

\newcommand*{\coloneqqrev}{= \mathrel{\vcenter{\baselineskip0.5ex \lineskiplimit0pt \hbox{\scriptsize.}\hbox{\scriptsize.}}}}
\makeatletter
\def\@email#1#2{%
 \endgroup
 \patchcmd{\titleblock@produce}
  {\frontmatter@RRAPformat}
  {\frontmatter@RRAPformat{\produce@RRAP{*#1\href{mailto:#2}{#2}}}\frontmatter@RRAPformat}
  {}{}
}%
\makeatother
\begin{document}

\preprint{AIP/123-QED}

\title[]{Holographic entanglement in spin network states: a focused review}
\author{Eugenia Colafranceschi}
\email{eugenia.colafranceschi@nottingham.ac.uk}
\author{Gerardo Adesso}%

\affiliation{School of Mathematical Sciences and Centre for the Mathematics and Theoretical Physics of Quantum Non-Equilibrium Systems, University of Nottingham, University Park Campus, Nottingham NG7 2RD, United Kingdom}

\date{\today}

\begin{abstract}
In the long-standing quest to reconcile gravity with quantum mechanics, profound connections have been unveiled between concepts traditionally pertaining to quantum information theory, such as entanglement, and constitutive features of gravity, like holography. Developing and promoting these connections from the conceptual to the operational level unlocks access to a powerful set of tools, which can be pivotal towards the formulation of a consistent theory of quantum gravity. Here, we review recent progress on the role and applications of quantum informational methods, in particular tensor networks, for quantum gravity models. We focus on spin network states dual to finite regions of space, represented as entanglement graphs in the group field theory approach to quantum gravity, and illustrate how techniques from random tensor networks can be exploited to investigate their holographic properties. In particular, spin network states can be interpreted as maps from bulk to boundary, whose holographic behaviour increases with the inhomogeneity of their geometric data (up to becoming proper quantum channels). The entanglement entropy of boundary states, which are obtained by feeding such maps with suitable bulk states, is then proved to follow a bulk area law, with corrections due to the entanglement of the bulk state. We further review how exceeding a certain threshold of bulk entanglement leads to the emergence of a black hole-like region, revealing intriguing perspectives for quantum cosmology.
\end{abstract}

\maketitle

\section{Introduction}

Holography has been a driving theme of research in quantum gravity since the discovery of the Bekenstein-Hawking area law for black hole entropy\cite{Bekenstein:1973ur,Hawking:1975vcx} and the discussion on information loss and Hawking radiation\cite{PhysRevD.14.2460,Page:1993wv}. 
Aspects and realisations of the holographic principle, originally proposed by 't~Hooft\cite{tHooft:1993dmi} and later developed by Susskind\cite{Susskind:1994vu} and Bousso \cite{Bousso:2002ju}, have been extensively studied at both classical and quantum level. Relevant instances include, out of a very wide range of contributions, early work on the microscopic interpretation of the black hole entropy\cite{Strominger:1996sh,Strominger:1997eq,Carlip:1998wz,Carlip:2002be}; on the recovering of gravitational dynamics from the thermodynamics of boundaries\cite{Padmanabhan:2013nxa,Padmanabhan:2015zmr}; on the 
duality between the gravitational theory of asymptotically anti de Sitter (AdS) spacetime and a conformal field theory (CFT) leaving on its boundary, known as AdS/CFT correspondence\cite{Maldacena:1997re,Witten:1998zw,Gubser:1998bc,Witten:1998qj} (and, within the latter, the Ryu-Takayanagi formula\cite{Ryu:2006bv,Ryu:2006ef} relating the boundary entanglement entropy to the area of a bulk surface); on holography in loop quantum gravity \cite{Markopoulou:1997hu,Krasnov:2009pd,Livine:2017xww}.

In recent years, an intriguing connection between gravity, holography and quantum entanglement has come to light. On one hand, several results point to entanglement as the \virg{glue} of spacetime\cite{VanRaamsdonk:2009ar,VanRaamsdonk:2010pw,Cao:2016mst}; on the other, entanglement turns out to be intimately tied to holography in quantum many-body systems\cite{Eisert:2008ur}, and quantum spacetime can indeed be understood, in several background-independent approaches to quantum gravity, as a collection of (fundamental, \virg{pre-geometric}) quantum entities\cite{penrose1972nature}, i.e.~as a (background-independent) quantum many-body system\cite{Oriti:2018dsg}.  Understanding the origin of the gravity/holography/entanglement threefold connection would therefore be a major step towards the formulation of a theory of quantum gravity\cite{Qi:2018ogs}. 

The main aim of this article is to review recent results\cite{Colafranceschi:2020ern,Colafranceschi:2021acz,Chirco:2021chk} that stand out for investigating  holography directly at the level of \textit{quantum gravity states}, in a \textit{quasi-local context} and via a \textit{quantum information language}. The focus is on finite regions of 3D quantum space modelled by spin networks, i.e.~graphs decorated by quantum geometric data (a formalism originally proposed by Penrose\cite{penrose1971angular}) which enter, as kinematical states, various background-independent approaches to quantum gravity\cite{Bodendorfer:2016uat,Perez:2004hj,Freidel:2005qe}. Crucially, such states are understood as arising from the entanglement of the quantum entities (\virg{atoms of space}) composing the spacetime microstructure, in the group field theory (GFT) framework\cite{Freidel:2005qe, Oriti:2011jm}; that is, as \textit{graphs of entanglement}. This formalism has the remarkable property of realising, directly at the level of the quantum microstructure of spacetime, the interrelation between entanglement and space connectivity supported by several results in quantum gravity contexts and beyond\cite{Ryu:2006bv,Ryu:2006ef,Swingle:2009bg,VanRaamsdonk:2009ar,VanRaamsdonk:2010pw,Cao:2016mst}. Moreover, as entanglement graphs, the spin network states are put in correspondence with tensor networks\cite{Orus:2013kga}, a quantum information language that efficiently encodes entanglement in quantum many-body systems. Such an information-theoretic perspective on spin network states is then exploited to investigate the role of entanglement (and quantum correlations more generally) in the holographic features of quantum spacetime, via tensor network techniques.

As further reviewed in this article, the aforementioned approach is shared by a rich body of work at the interface of quantum gravity, quantum information and condensed matter physics which looked at entanglement on spin networks as a tool for probing and reconstructing geometry. It includes the modelling of quantum black holes and the computation of the horizon entropy\cite{Livine:2005mw,Livine:2007sy,Oriti:2015rwa,Oriti:2018qty}; the reconstruction of a notion of distance on spin networks from entanglement\cite{Livine:2006xk,Feller:2015yta}; the characterisation of the entanglement entropy between an arbitrary region of a spin network and its complement\cite{Donnelly:2008vx,Donnelly:2011hn,Delcamp:2016eya,Livine:2017fgq}; the use of entanglement to glue quantum polyhedra dual to spin network vertices\cite{Baytas:2018wjd,Bianchi:2018fmq}; the study of the holographic properties of spin network states\cite{Bianchi:2018fmq,Feller:2017jqx,Chirco:2017vhs}.

The results of Ref.~\onlinecite{Colafranceschi:2020ern,Colafranceschi:2021acz,Chirco:2021chk} reviewed here investigate holography in finite regions of quantum space from two different perspectives: (i) by studying the flow of information from the bulk to the boundary, and (ii) by analysing the information content of the boundary, and its relationship with the bulk. The idea behind perspective (i) is the possibility (pointed out for the first time in Ref.~\onlinecite{Chen:2021vrc}) to interpret every spin network state as a bulk-to-boundary map, and the holographic character of the latter is traced back to how close it comes to being an isometry. The impact of combinatorial structure and geometric data of spin network states (matching random tensor networks) on the \virg{isometry degree} of the corresponding bulk-to-boundary maps is then studied, by relying on random tensor network methods. Perspective (ii) focuses on the entanglement entropy content of boundary states, obtained by feeding the aforementioned bulk-to-boundary maps with a bulk input state, upon varying the latter. The result is twofold: on one hand, a bulk area law for the boundary entropy, with corrections due to the bulk entanglement; on the other, the emergence of horizon-like surfaces when increasing the entanglement content of the bulk.

This focused review is structured in four sections. The first one is dedicated to the quantum gravity framework: subsections \ref{quantumofspace} and \ref{spinnet} show the logical path from a quantised, elementary portion of space (a tetrahedron) to extended discrete quantum geometries, and the dual spin network description; subsection \ref{GFT} presents group field theories, quantum gravity models in which spin networks can be readily understood as graphs of entanglement, and as kinematic quantum gravity states; finally, subsection \ref{TNper} illustrates the tensor network perspective on spin network states. In section \ref{literature} we give an overview of earlier results on the study of entanglement on spin network states and its role in reconstructing geometry. Section \ref{partII} is dedicated to random tensor network techniques, adapted to the considered quantum gravity framework; more specifically, it shows how to compute the Rényi-2 entropy of a certain class of spin network states via a statistical model. Section \ref{partIII} contains the aforementioned results on the holographic features of spin network states matching random tensor networks, from the perspective of bulk-to-boundary maps (subsection \ref{btob}) and of the entanglement entropy of boundary states (subsection \ref{bstates}).

\section{Quantum gravity states as entanglement graphs}\label{partI}

Several approaches to quantum gravity, e.g. loop quantum gravity\cite{Bodendorfer:2016uat}, spinfoam models\cite{Perez:2004hj} and group field theories\cite{Freidel:2005qe, Oriti:2011jm} (GFT), describe regions of 3D space via \textit{spin networks}, graphs decorated by quantum geometric data. We review how spin networks can be constructed from elementary portions of space (e.g. small tetrahedra) quantised and glued together to form extended (discrete) spatial geometries; crucially, the gluing derives from entanglement, and spin networks can thus be regarded as the entanglement structure of many-body states for the set of elementary tetrahedra. We then introduce group field theories\cite{Freidel:2005qe, Oriti:2011jm}, quantum gravity models where the above picture is realised, and spin networks from many-body entanglement can be understood as kinematical quantum gravity states. We conclude by reviewing recent results\citep{Colafranceschi:2020ern} on the formal correspondence between spin network states and tensor networks.

\subsection{Quantum tetrahedron and the dual spin network vertex}\label{quantumofspace}

Consider an elementary portion of 3D space, a tetrahedron, whose faces are labelled by an index $i=1,2,3,4$. The (classical) geometry of the tetrahedron can be described by four vectors $\{\vec{L}_i\}_{i=1}^4$, with $\vec{L}_i$ normal to the $i$-th face and having length equal to the face area, which satisfy the \textit{closure constraint}\cite{rovelli_vidotto_2014}:
\begin{equation}
\sum_{i=1}^4 \vec{L}_i=0.
\end{equation}
The equivalence class of the four vectors $\{\vec{L}_i\}_{i=1}^4$ under global rotations encodes a geometrical configuration of the tetrahedron. Note that, as the vectors $\{\vec{L}_i\}_{i=1}^4$ are elements of the $su(2)$ Lie algebra, we can equivalently describe the geometry of the tetrahedron via the dual $SU(2)$ group elements $\{g^i\}_{i=1}^4$ (more precisely, via the equivalence class of $\{g^i\}_{i=1}^4$ under global $SU(2)$ action). 

The quantisation of the phase space of geometries of a tetrahedron\cite{Barbieri:1997ks,Freidel:2010xna,Livine:2013tsa} leads to the Hilbert space $\mathcal{H}=L^2(G^4/G)$, where $G=SU(2)$; i.e.~the quantum state of geometry of a tetrahedron is described by a wave-function $f(\vec{g})$, where $\vec{g}=\{g^1,g^2,g^3,g^4\}$, that satisfies
\begin{equation}\label{inv}
f(\vec{g})=f(h\vec{g})\quad \forall h \in SU(2),
\end{equation} 
with $h\vec{g}\coloneqq \{ hg^1,hg^2,hg^3,hg^4\}$. 

By the Peter-Weyl theorem, the wave-function $f(\vec{g})$ can be decomposed into irreducible representations $j\in \frac{\mathbb{N}}{2}$ of $SU(2)$\cite{Martin-Dussaud:2019ypf}:
\begin{equation}\label{PW}
f(\vec{g})=\sum_{\vec{j}\vec{m}\vec{n}} f^{\vec{j}}_{\vec{m}\vec{n}}\prod_{i=1}^{4}\sqrt{2j^i+1}D^{j^i}_{m^in^i}(g^i)
\end{equation}
where we used a vector notation for set of variables attached to the four faces of the tetrahedron, e.g. $\vec{j}=\{j^1,j^2,j^3,j^4\}$; the \textit{magnetic index} $m^i$ ($n^i$) labels a basis of the $j^i$-representation space $V^{j^i}$ (its dual ${V^{j^i}}^*$); and $D^{j^i}_{m^in^i}(g^i)$ is the Wigner matrix representing the group element $g^i$. When taking into account the gauge symmetry (see Eq.~\eqref{inv}), both the expansion coefficients and the Wigner matrices end up contracted with a $SU(2)$-invariant tensor, i.e.~an \textit{intertwiner} $\iota$, pertaining to the Hilbert space
\begin{equation}\label{int}
\mathcal{I}^{\vec{j}}\coloneqq \text{Inv}_{SU(2)}\left[V^{j^1}\otimes ... \otimes V^{j^4}\right]
\end{equation}
and ensuring the gauge invariant recoupling of the four spins $\{j^i\}_{i=1}^4$. Equation \eqref{PW} then becomes\cite{Martin-Dussaud:2019ypf}
\begin{equation}
f(\vec{g})=\sum_{\vec{j}\vec{n}\iota} f^{\vec{j}}_{\vec{n}\iota}s^{\vec{j}}_{\vec{n}\iota}(\vec{g}),
\end{equation}
where 
\begin{equation}\label{spinwf}
s^{\vec{j}}_{\vec{n},\iota}(\vec{g})\coloneqq \iota_{\vec{m}}\prod_{i=1}^4\sqrt{2j^i + 1}\wig{j^i}{m^in^i}{g^i}
\end{equation}
is the generic element of the \textit{spin network basis}, and $f^{\vec{j}}_{\vec{n}\iota}\coloneqq f^{\vec{j}}_{\vec{m}\vec{n}} \iota_{\vec{m}}$. We denote by $d_j\coloneqq 2j+1$ the dimension of the representation space $V^j$, and by $D_{\vec{j}}$ the dimension of the intertwiner space $\mathcal{I}^{\vec{j}}$.

The spin network basis $\{|\vec{j}\vec{n}\iota \rangle\}$ diagonalises the area and volume operators\cite{Rovelli:1994ge,Rovelli:1995ac,Ashtekar:1996eg,Ashtekar:1997fb}, and thus possesses a clear geometrical interpretation; more specifically, the $SU(2)$ spin $j^i$ determines the area of the $i$-face of the tetrahedron, while the intertwiner $\iota$ determines its volume. 

\begin{figure}[t]
	\centering
	\includegraphics[width=0.5\linewidth]{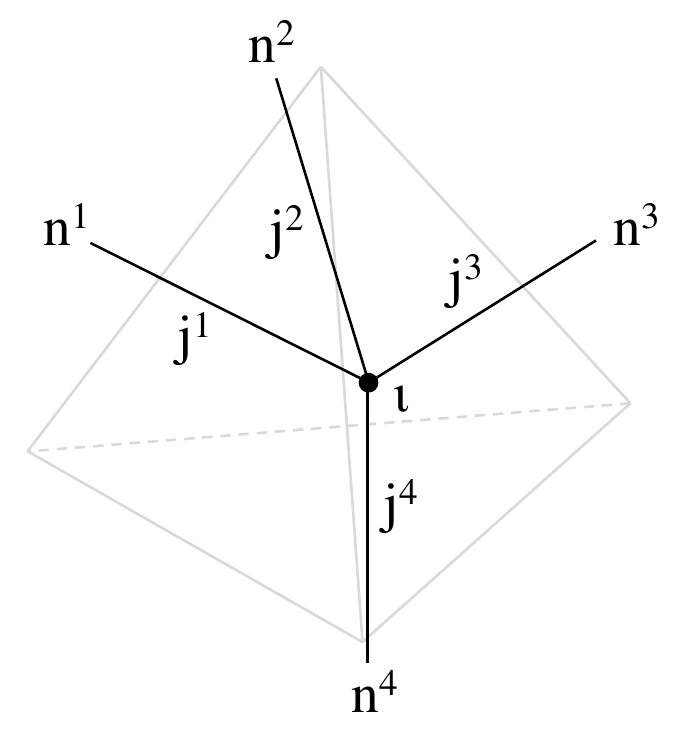}\caption{Spin network vertex (black) representing the tetrahedron (grey). Every edge $e^i$ of the vertex is dual to a face of the tetrahedron, carries a representation spin $j^i$ and, at the free endpoint, the magnetic index (spin projection) $n^i$; the intertwiner $\iota$ deriving from the recoupling of the four spins is associated to the intersection points of the four edges. }
	\label{vertex_fig}
\end{figure}

The quantum tetrahedron can be graphically represented as a vertex with four \textit{edges}, each one identified by a \textit{colour} $i$, where the $i$-th edge (denoted by $e^i$) is dual to the $i$-th face of the tetrahedron and carries the corresponding quantum data (see figure \ref{vertex_fig}): in the group basis, the edge $e^i$ carries a group variable $g^i$; in the spin network basis, the edge $e^i$ carries a spin $j^i$ and, at the free endpoint, the magnetic index $n^i$, while the intertwiner quantum number $\iota$ is attached to the vertex itself. This structure is called \textit{spin network vertex}. 

At the level of the Hilbert space of the quantum tetrahedron, the spin network decomposition performed via the Peter-Weyl theorem reads 
\begin{equation}
\h = L^2(G^4/G)=\bigoplus_{\vec{j}}\left(\mathcal{I}^{\vec{j}}\otimes \bigotimes_{i=1}^4V^{j^i}\right),
\end{equation}
where the intertwiner space $\mathcal{I}^{\vec{j}}$ is defined in Eq.~\eqref{int}. 

The above construction can be easily generalised to any elementary polyhedron. In particular, the quantum version of a $(d-1)$-simplex (which is the simplest possible $(d-1)$-polytope) is dual to a $d$-valent vertex and described by the Hilbert space 
\begin{equation}\label{dHilbert}
\h=L^2(G^d/G)=\bigoplus_{\vec{j}}\left(\mathcal{I}^{\vec{j}}\otimes \bigotimes_{i=1}^dV^{j^i}\right).
\end{equation} 
In the following we take into account this generalisation and adopt, for the $\vec{j}$-spin sector, the notation 
\begin{equation}\label{hj}
\h_{\vec{j}}\coloneqq \mathcal{I}^{\vec{j}}\otimes \bigotimes_{i=1}^dV^{j^i}.
\end{equation}
Also, to clarify the role of the different degrees of freedom of a spin network vertices, for some equations we write the basis element $\ket{\vec{j}\vec{n}\iota}$ of $\h_{\vec{j}}$ in the form 
\begin{equation}\label{baisdec}
\ket{\vec{j}\vec{n}\iota}= \ket{j^1n^1}...\ket{j^dn^d} \ket{\vec{j}\iota},
\end{equation} 
i.e.~as explicit tensor product of the basis states of the intertwiner and representation spaces: $\ket{\vec{j}\iota}\in \mathcal{I}^{\vec{j}}$ and $\ket{j^in^i} \in V^{j^i}$, respectively.
\subsection{Gluing tetrahedra: spin networks for 3D quantum geometries}\label{spinnet}

\begin{figure}[t]
	\centering
\includegraphics[width=0.9\linewidth]{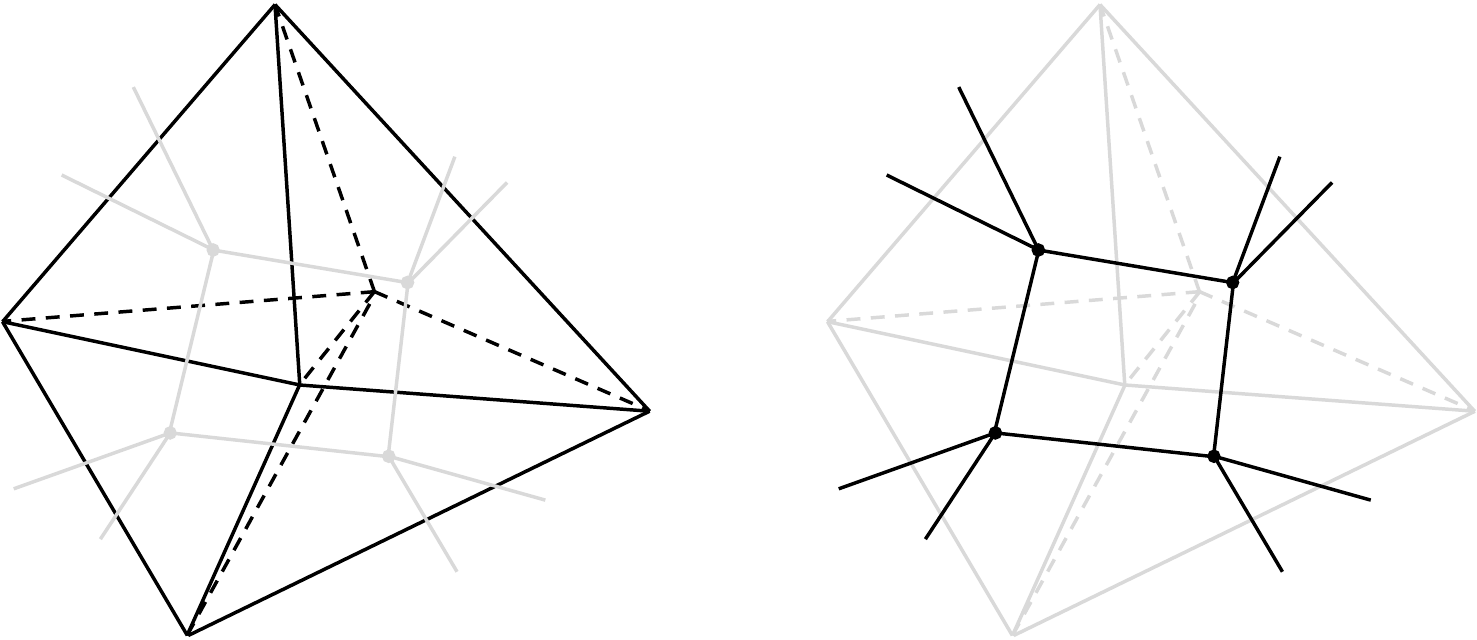}\caption{The quantum geometry of a simplicial complex (highlighted in black on the left) is described by the dual spin network graph (highlighted in black on the right). Adapted with permission from Chirco \textit{et al.}, Phys. Rev. D \textbf{105}(4), 046018 (2022).}
	\label{graph_fig}
\end{figure}

A region of 3D space can be arbitrary well approximated by a collection of (suitably small) polyhedra adjacent to each other. As we are going to show, the quantum geometry of such a discrete space can be described by a set of interconnected spin network vertices corresponding to the single polyhedra\cite{Oriti:2013aqa}; the result is a \textit{spin network graph}\cite{Perez:2004hj}, i.e a graph $\gamma$ dual to the space partition and decorated by quantum geometric data, as showed in figure  \ref{graph_fig}. 

More precisely, a spin network graph represents the quantum version of a \textit{twisted geometry}\cite{Freidel:2010aq,Dittrich:2010ey,Freidel:2018pvm}. The latter is a collection of polyhedra in which adjacent faces possess the same area but have, in general, different shape and/or orientation. That is, only  \virg{neighbouring relations} are present in a twisted geometry: the planes of adjacent faces are not necessarily parallel. Twisted geometries thus differ from standard Regge triangulations in which faces of neighbouring polyhedra, having the same area, shape and orientation, perfectly adhere to each other.  In the following, we explain how spin networks can arise from the gluing of vertices dual to polyhedra, where by \virg{gluing} we mean establishing an adjacency relationship between them as defined in twisted geometries.

Consider a set $v=1,...,N$ of open spin network vertices of valence $d$, which is described by the Hilbert space $\h_N=L^2\left(G^{d\times N}/G^N\right)$.  We illustrate the gluing of vertices with an example. Given two vertices $v$ and $w$, we want to glue the $i$-th edge of $v$ (denoted by $e_v^i$), which carries the group variable $g_v^i$, with the $j$-th edge of $w$ (denoted by $e_w^j$), which carries $g_w^j$. As both edges are outgoing, the resulting link from $v$ to $w$ (denoted by $\ell_{vw}^{ij}$) carries the group element $g_v^i (g^j_w)^{-1}$. Once connected, the two vertices are thus invariant under the simultaneous right action of the group on the edges $e_v^i$ and $e_w^j$, as $g_v^i h (g^j_wh)^{-1}=g_v^i (g^j_w)^{-1}$ $\forall h\in SU(2)$. Starting from the set of open vertices in the state $\psi\in \h_N$, such a symmetry (that is, the gluing of edges) can be implemented via the following group averaging\cite{Livine:2011yb,Oriti:2013aqa}:
\begin{equation}\begin{split}\label{conv}
\int \diff h\psi(...,g^i_v h, ..., g^j_w h,...)=&\psi_\gamma(...,g_v^i (g^j_w)^{-1},...),
\end{split}
\end{equation}
which in fact causes the resulting $\psi_\gamma$ to depend on $g_v^i$ and $g_w^j$ only through the product $g_v^i (g^j_w)^{-1}$. The wave-function $\psi_\gamma$ is then associated to a graph $\gamma$ involving the link $\ell_{vw}^{ij}$. In the group basis, the geometric data attached to a spin network graph thus consist in a group element on every edge of the graph, with gauge invariance at each vertex. This structure is therefore described by the Hilbert space $\h_\gamma=L^2(G^L/G^N)$, where $L$ is the number of links of $\gamma$\cite{Perez:2004hj}.

\begin{figure}[t]
\centering 
\subfigure[\label{fig:hred}]{\includegraphics[width=0.95\linewidth]{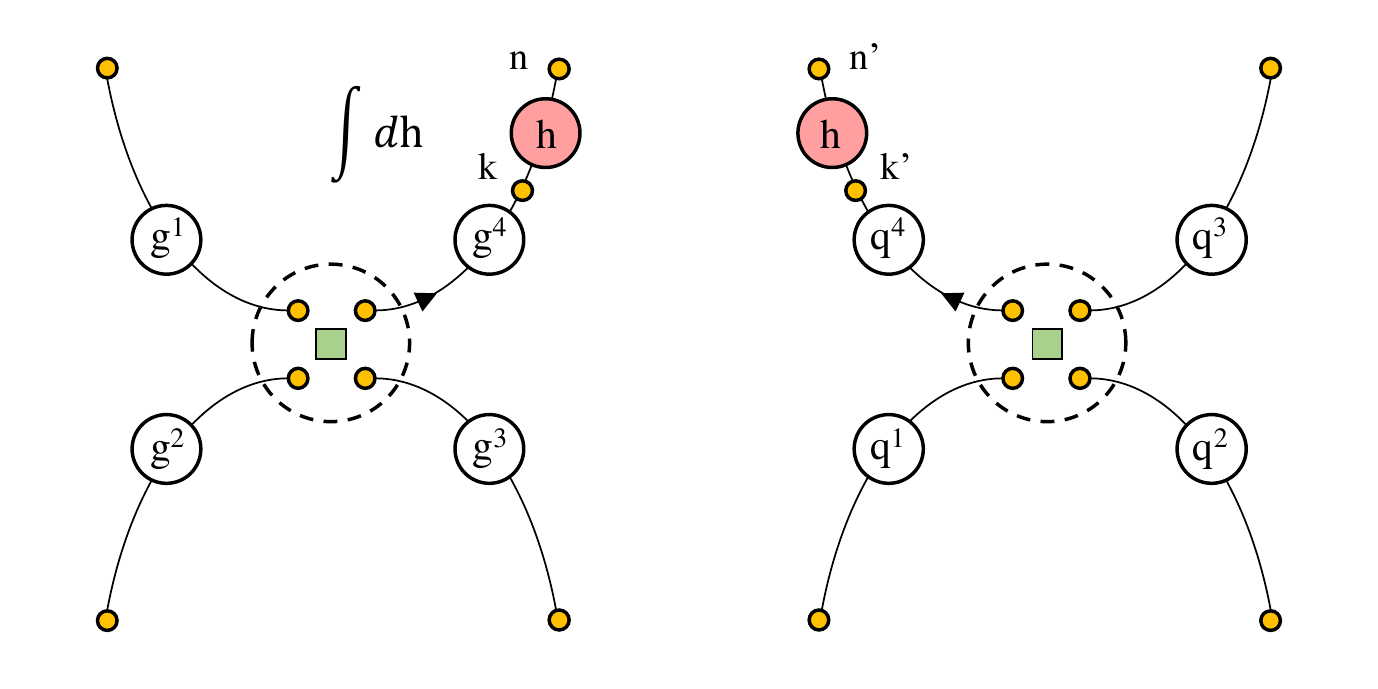}
}
\centering
\subfigure[\label{fig:green}]{\includegraphics[width=0.95\linewidth]{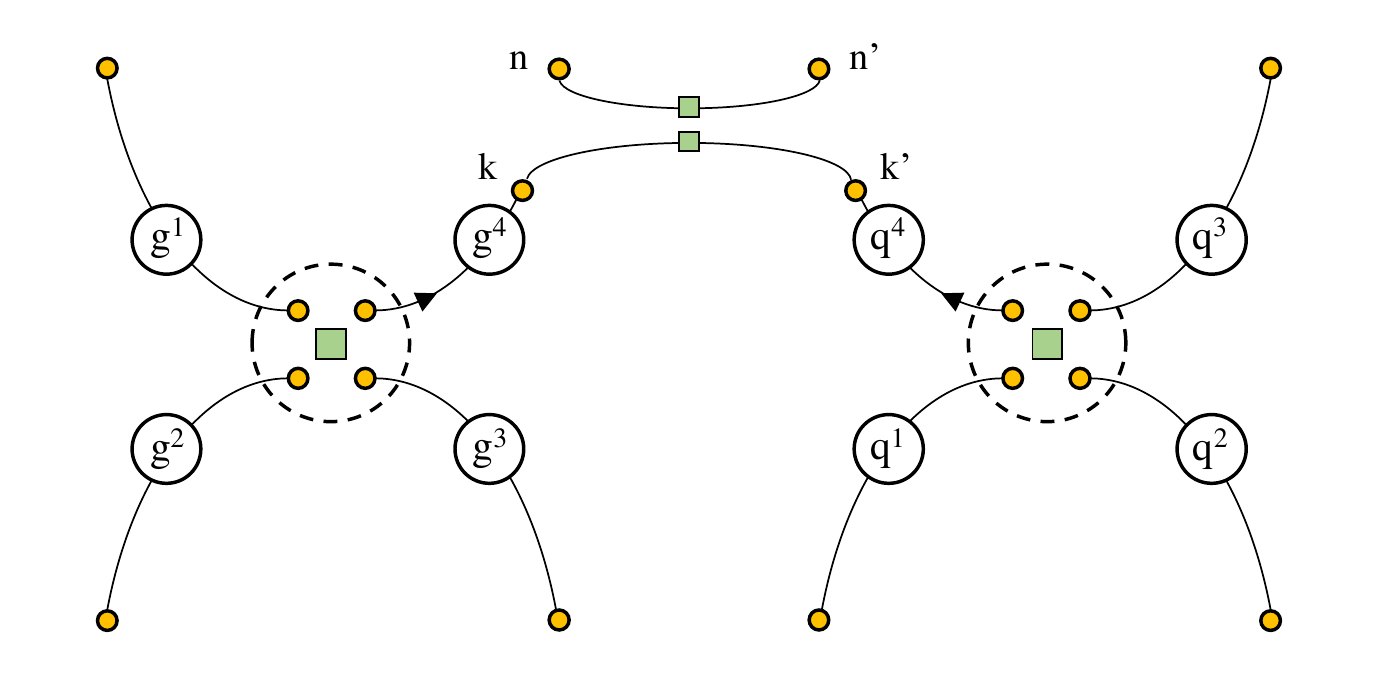}
}
\caption{Gluing of two spin network vertices performed by acting on two open edges with the same group element and integrating over the latter. Group variables are depicted as large white disks (except for the element $h$ through which the group acts, highlighted in red) and magnetic indices as small yellow disks; intertwiner tensors are instead represented by green squares. The group averaging is depicted in panel (a) and returns a pair of bivalent intertwiners contracting the magnetic indices of the two edges, as shown in panel (b). }
\label{fig:integral}
\end{figure}

\subsubsection{Spin networks as entanglement graphs} 

In the spin network basis, gluing edges corresponds to entangling the degrees of freedom attached to their free ends\cite{Colafranceschi:2020ern}. We clarify this point with the following example.
Consider the gluing of two four-valent spin network vertices described by the wavefunction $\psi$:
\begin{multline}\label{glue}
\int dh \psi(g^1,\dots,g^4h,q^1,\dots,q^4h)\\= \psi_{\vec{j}\vec{j}' \vec{n}\vec{n}'\iota\iota'}\int dh s^{\vec{j}}_{\vec{n},\iota}(g^1,\dots,g^4h) s^{\vec{j}'}_{\vec{n}',\iota'}(q^1,\dots,q^4h)
\end{multline}
The integral of the spin network basis elements, which is the factor implementing the gluing, is represented in figure \ref{fig:integral} and performed in the following. To simplify the notation, the label $4$ is removed from all quantum numbers (e.g. $j^4$ is denoted just as $j$); we also adopt the notation $n^{123}=\lbrace n^1,n^2,n^3\rbrace$. By substituting to Eq.~\eqref{glue} the expression of Eq.~\eqref{spinwf} one obtains
\begin{multline}\label{3a}
\int dh \psi(g^1,\dots,g^4h,q^1,\dots,q^4h)\\= \psi_{\vec{j}\vec{j}' \vec{n}\vec{n}'\iota\iota'}s^{\vec{j}}_{n^{123}k,\iota}(\vec{g}) s^{\vec{j}'}_{n'^{123}k',\iota'}(\vec{q})\int dh \wig{j}{kn}{h}\wig{j'}{k'n'}{h}
\end{multline}
The integral of the Wigner matrices, sketched in figure \ref{fig:hred}, is well-known in representation theory and yields
\begin{equation}\label{wig}
 \int dh \wig{j}{kn}{h}\wig{j'}{k'n'}{h}=\delta_{jj'}I_{kk'}I_{nn'}
\end{equation}
where $I_{kk'}$ is a bivalent intertwiner in the space $V^{j}\otimes V^{j}$ attached to the free ends of the to-be-glued edges:
\begin{equation}\label{bivalent}
I_{kk'}\coloneqq \frac{(-1)^{j+k}}{\sqrt{2j+1}}\delta_{k,-k'}
\end{equation}
By inserting Eq.~\eqref{wig} into Eq.~\eqref{3a} the latter becomes (see figure \ref{fig:green})
\begin{multline}\label{glued}
\int dh \psi(g^1,\dots,g^4h,q^1,\dots,q^4h)\\= \left(\psi_{\vec{j}\vec{j}' \vec{n}\vec{n}'\iota\iota'}\delta_{jj'}I_{nn'}\right) s^{\underline{j}j}_{n^{123}k,\iota}(\vec{g}) s^{\underline{j}'j}_{n'^{123}k',\iota'}(\vec{q})I_{kk'}
\end{multline}
In the expression above, both the state coefficient and the spin network basis elements are contracted with a bivalent intertwiner. Crucially, this is equivalent to projecting $\ket{\psi}$ on the following state of $V^{j}\otimes V^{j}$:
\begin{equation}\label{singlet}
\ket{\ell}=\sum_{kk'}I_{kk'}\ket{k}\ket{k'}=\sum_{k}\frac{(-1)^{j+k}}{\sqrt{2j+1}}\ket{k}\ket{-k} 
\end{equation}
which is a singlet state. The entanglement between the two edges composing the link can be quantified via the von Neumann entropy. Denoting by $\rho_s$ and $\rho_t$ the reduced density matrices of the edges attached to the source and target vertex of the link $\ell$, respectively, one can easily check that
\begin{equation}
S(\rho_s)=S(\rho_t)=\log d_j
\end{equation}
i.e.~the entanglement entropy of the two subsystems reaches its maximum possible value: the state $\ket{\ell}$ is maximally entangled.

Therefore, starting from a set of open spin network vertices, the gluing of pairs of their edges is performed by entangling, in a singlet states, the spins on the corresponding free ends. \textit{The connectivity pattern of a set of vertices can thus be understood as an entanglement pattern among the degrees of freedom attached to the free ends of their open edges}. 

Finally note that, in the spin network basis, the graph resulting from the gluing of spin network vertices is decorated as follows: every link $\ell$ carries a spin $j_\ell$, and every vertex $v$ carries an intertwiner $\iota_v$. Open edges, when present, carry an additional quantum number: the (non-contracted) magnetic index at their free end (which causes the spin network to transform non-trivially under gauge transformations acting on its boundary).

\subsubsection{Constructing entanglement graphs of arbitrary connectivity}
 
The construction of spin network states of arbitrary connectivity $\gamma$ from many-body states associated to $N$ open vertices (where $N$ is the number of vertices in $\gamma$) has been rigorously defined in Ref.~\onlinecite{Colafranceschi:2020ern}. The first ingredient is a description of the combinatorial structure of graphs in terms of individual coloured vertices. In graph theory, the connectivity pattern of a set of $N$ vertices (whose edges are not distinguished by a colour) is encoded in the \textit{adjacency matrix}, i.e.~a $N \times N$ symmetric matrix $A$ defined as follows: the generic element $A_{xy}$ takes value 1 if vertices $x$ and $y$ are connected, and 0 otherwise. This encoding can be easily generalised to the case in which edges departing from vertices are distinguished by a colour $i$, as it happens with spin network vertices. Assuming the absence of 1-vertex loops, the generalised adjacency matrix takes the form
\begin{equation}\label{A}
A=\left(\begin{array}{cccc}
0_{d\times d} &A_{12} &\dots&A_{1N}\\
 & 0_{d\times d}&\\
 &  &  \ddots \\
 &  &  & 0_{d\times d}
\end{array}\right)
\end{equation} 
where $A_{vw}$ is now a $d\times d$ matrix (and $0_{d\times d}$ stands for the null $d\times d$ matrix), with element $\left(A_{vw}\right)_{ij}$ equal to 1 if vertices $v$ and $w$ are connected along edges of colour $i$ and $j$, respectively (i.e.~$e_v^i$ and $e_w^j$ are glued together), and 0 otherwise. To simplify the notation, and since the edge colouring does not play any particular role, one usually assumes that vertices can be connected only along edges of the same colour. The matrix $A_{vw}$ then takes a diagonal form:
\begin{equation}
A_{vw}=\left(\begin{array}{cccc}\label{Avw}
    a^1_{vw} & 0 & \dots & 0\\
    0 & \ddots &   &\\
    \vdots &  & \ddots &\\
    0 &  &  & a^d_{vw}
  \end{array}\right)
\end{equation}
with $a^i_{vw}$ equal to 1 (0) if vertices $v$ and $w$ are connected (not connected) along their edges of colour $i$; a link formed by $e_v^i$ and $e_w^i$ is denoted as $\ell_{vw}^i$.

The generalised adjacency matrix defined by Eqs. \eqref{A} and \eqref{Avw} thus encodes the connectivity pattern $\gamma$ of a set of $N$ vertices; that is, \virg{who is glued to whom}. The next ingredient for the implementation of $\gamma$ on a set of open vertices is the operator performing the gluing of edges, defined as follows. The operator $\mathbb{P}_{\ell_{vw}^i}$ creating the link $\ell_{vw}^i$ acts on the edges $e_v^i$ and $e_w^i$ by projecting their state onto the subspace characterised by the gluing symmetry (invariance under simultaneous right action of the group):
	\begin{equation}
	\mathbb{P}_{\ell_{vw}^i}\coloneqq\int \diff \text{h}\diff g_v^i\diff g_w^i ~ ~\ket{g_v^i h}\bra{g_v^i}\otimes \ket{g_w^i h}\bra{g_w^i}.
	\end{equation}
A spin network state associated to the generic graph $\gamma$ can then be obtained from a set of open vertices in the state $\psi\in \h_N$ by applying to the latter a set of gluing operators according to the adjacency matrix $A$ of $\gamma$:
\begin{equation}\label{P}
\ket{\psi_\gamma}=\left(\bigotimes_{a^i_{vw}=1}\mathbb{P}_{\ell_{vw}^i}\right) \ket{\psi}.
\end{equation}
As follows from Eq.~\eqref{glued}, in the spin network basis the gluing operator is a projection of edge spins onto maximally entangled states. The graph $\gamma$ of the spin network state of Eq.~\eqref{P} is thus realised as a \textit{pattern of entanglement} of a set of vertices. Spin networks regarded as arising from the entanglement structure of states describing a collection of spin network vertices are also referred to as \textit{entanglement graphs}.
\subsection{Group field theories}\label{GFT}

A group field theory\cite{Freidel:2005qe, Oriti:2011jm} (GFT) is a theory of a quantum field $\phi$ defined on $d$ copies of a group manifold $G$.
In the GFT model of simplicial quantum gravity $\phi$ is a bosonic field whose fundamental excitation is an elementary polyhedron, specifically the $(d-1)$-simplex dual to the $d$-valent spin network vertex introduced in section \ref{quantumofspace}. The action of the model takes the following form:
\begin{multline}
S_d[\phi]=\int \diff \vec{g} \diff \vec{q}\phi(\vec{g})\mathcal{K}(g^i\left(q^i)^{-1}\right) \phi(\vec{q}) \\+ \frac{\lambda}{d+1}\int \prod_{i\neq j=1}^{d+1}\diff g_i^j\mathcal{V}(g_i^j (g_j^i)^{-1})\phi(\vec{g_1})...\phi(\vec{g}_{d+1})
\end{multline}
where $\vec{g}=\{g^1,...,g^d\}$; $\mathcal{K}(g^i\left(q^i)^{-1}\right)$ is the kinetic kernel, responsible for the gluing of polyhedra (spin network vertices) which gives rise to extended spatial geometries (spin network graphs); $\lambda$ is a coupling constant and $\mathcal{V}(g_i^j (g_j^i)^{-1})$ is the interaction kernel, which determines the interaction processes of polyhedra that generate $d$-dimensional spacetime manifolds of arbitrary topology. In particular, due to the simplicial interpretation of field quanta, the Feynman amplitudes of the theory
are given by simplicial path integrals (a characteristic shared with simplicial approaches to quantum
gravity\cite{AMBJORN199242}) or, equivalently, spin foam models\cite{Perez:2004hj} (representing \virg{histories} of spin networks).

The GFT Fock space is constructed from the Hilbert space $\h$ of the $(d-1)$-simplex (equivalently, the dual $d$-valent vertex) defined in Eq.~\eqref{dHilbert}:
\begin{equation}
    \mathcal{F}(\h)=\bigoplus_N \mathrm{sym}\left( \underbrace{\h\otimes ... \otimes \h}_N\right).
\end{equation}
It includes the spin network states in the form of Eq.~\eqref{P}, symmetrised over the vertex labels. Crucially, the symmetry under relabelling of vertices can be understood as a discrete version of diffeomorphism invariance\cite{Colafranceschi:2020ern} (which is a necessary condition for background independence), as the vertex labels behave like \virg{coordinates} over the spatial manifold described by the spin network. 

Let us finally remark that spin networks arise, in this context, from the entanglement properties of many-body states describing a set of (indistinguishable) spin network vertices. More specifically, the entanglement structure of the many-body state can be identified with the graph formed by the vertices. In the following, we present the correspondence between spin network states and tensor networks, a quantum information language that realises an analogous graphical encoding of many-body entanglement.

\subsection{The tensor network perspective}\label{TNper}

Consider a many-body system composed of $N$ $d$-dimensional spins $s_1,...,s_N$. A generic state for the system,
\begin{equation}
\ket{\Psi}=\sum_{s_1...s_N}C_{s_1...s_N}\ket{s_1...s_N}
\end{equation} 
is described by $d^N$ complex coefficients $C_{s_1...s_N}$. The computational cost of this description can however be reduced by considering a \textit{tensor network decomposition}\cite{Orus:2013kga} of the state. It consists in replacing the tensor $C_{s_1...s_N}$ with a collection of smaller tensors $T^{s_i}_i$ interconnected via auxiliary indices $\vec{a}_i=a_i^1,...,a_i^r$ (for simplicity, we assume each one having dimension $D$): 
\begin{equation}\begin{split}
C_{s_1...s_N}&=\Tr_{\mathcal{N}} \left[T^{s_1}_{1}...T^{s_N}_{N}\right] \\&=\left(T_1^{s_1}\right)^{\vec{a_1}}...\left(T^{s_N}_N\right)^{\vec{a_N}} \prod_{(A_{vw})_{ij}=1}\delta_{a_p^ia_q^j}
\end{split}
\end{equation}
where $\Tr_{\mathcal{N}}$ symbolises the trace over the auxiliary indices performed according to a combinatorial pattern $\mathcal{N}$ of the physical spins, $A$ is the adjacency matrix describing the network $\mathcal{N}$ and repeated indices are summed over. Note that the number of parameters needed to describe the tensor network has a polynomial scaling in the system size $N$, instead of a exponential one\cite{Orus:2013kga}; in the case we considered, it is given by $NdD^r$.

Spin networks regarded as entanglement graphs (according to the discussion of section \ref{spinnet}) formally correspond~\cite{Colafranceschi:2020ern} to a particular class of tensor networks, called \textit{projected entangled pair states}\cite{Verstraete:2004cf,Cirac_2009} (PEPS). A PEPS is a collection of maximally entangled states $\ket{\phi}=\sum_{a=1}^D \ket{a}\ket{a}$ of pairs of auxiliary systems projected locally onto physical systems $s_1,...,s_N$, with the entangled pairs corresponding to the \textit{links} of the resulting network $\mathcal{N}$. Let $\ket{\phi_\ell}$ be the maximally entangled state corresponding to link $\ell$ of $\mathcal{N}$, and let $Q_i$ be the operator at site $i$ projecting the auxiliary systems onto the physical one $s_i$; then
\begin{multline}
\ket{\Psi}=Q_1 \otimes Q_2\otimes ... \otimes Q_N \bigotimes_\ell \ket{\phi_\ell}\\=\sum_{s_1...s_N} \Tr_{\mathcal{N}} \left[T_{s_1}^{1}...T_{s_N}^N\right] \ket{s_1...s_N}
\end{multline} 
where the tensor $T^{s_i}_{i}$ has elements $\left(T^{s_i}_i\right)^{a_i^1a_i^2...}=\bra{s_i}Q_i\ket{a_i^1a_i^2...}$. The network $\mathcal{N}$ thus corresponds to the \textit{pattern of entanglement} of the physical spins $s_1,...,s_N$; in particular, the connectivity of $\mathcal{N}$ is realised by pairs of auxiliary degrees of freedom in a maximally entangled state.

Similarly, spin networks can be understood as arising from the entanglement structure of a many-body system, as explained in section \ref{spinnet}. The degrees of freedom encoding the connectivity of the spin network are the ones living in the representation spaces attached to the edge free-ends. Specifically, a pair of edges $e_v^i$ and $e_w^i$ is glued into a link $\ell_{vw}^i$ when the spins living on $V^{j_v^i}$ and $V^{j_w^i}$ are in a singlet states. The spin network counterpart to the link state $\ket{\phi_\ell}$ is thus the one of Eq.~\eqref{singlet}, that we rewrite here for clarity:
\begin{equation}\label{link}
\ket{\ell_{vw}^i}\coloneqq \frac{1}{\sqrt{d_{j}}} \sum_{k}(-1)^{j+k}\ket{k}\ket{-k} \in V^{j_v^i}\otimes V^{j_w^i}
\end{equation}
where $j_v^i=j_w^i=j$. 

Therefore, tensor networks and (completely generic) spin networks have in common the interpretation of links of the graph/network as maximally entangled pairs of systems (auxiliary degrees of freedom for the first, edge-spins for the latter). However, the spin network wave-function $\psi_\gamma$ is not, in general, a tensor network; that is, it does not necessarily factorise over single-vertex tensors. 

Nevertheless, spin network states obtained from the gluing of open vertices in the factorised state
\begin{equation}
\psi^{\vec{j}_1... \vec{j}_N}_{\vec{n}_1...\vec{n}_N\iota_1...\iota_N} = (f_1)^{\vec{j}_1}_{\vec{n}_1 \iota_1}... (f_N)^{\vec{j}_N}_{\vec{n}_N \iota_N}
\end{equation}
do formally correspond to tensor networks. In particular, they can be understood as PEPS, as the gluing procedure is effectively a projection of link states onto single-vertex states:
\begin{equation}\label{tnstate}
   |\psi_\gamma \rangle = \left(\bigotimes_{\ell \in \gamma}\langle \ell| \right) \bigotimes_v |f_v\rangle 
\end{equation}
where $\ell$ is a short notation for the generic link $\ell_{vw}^i$, $\ket{\ell}$ is the link state defined in Eq.~\eqref{link} and $f \in \h_{\vec{j}}$ is a fixed-spins vertex state. 

Note that, when regarding Eq.~\eqref{tnstate} as a tensor network, the spins $j$ on the graph $\gamma$ correspond to \virg{bond dimensions} of the tensor network indices. However, in the spin network formalism the spins are not fixed parameters (as are the tensor network bond dimensions), but \textit{dynamical variables}. Therefore, only the \virg{fixed-spins case} given by Eq.~\eqref{tnstate} formally corresponds to an ordinary tensor network. The generalised case with link and vertex wave-functions spreading over all possible spins thus qualifies as a superposition of tensor networks. Furthermore, given the bosonic nature of the discrete entities the individual tensors are associated to, spin networks obtained from factorised many-body states correspond to (superpositions of) \textit{symmetric} tensor networks (for more details, see Ref.~\onlinecite{Colafranceschi:2020ern}).

\section{Entanglement and correlations on spin networks to probe and reconstruct geometry} \label{literature}

In this section we give an overview of a series of works on the study of correlations and entanglement entropy on spin networks which, in the spirit of the results presented in part \ref{partIII}, are based on the interplay between quantum gravity and quantum information and/or condensed matter physics. The results are grouped by theme and presented in mainly chronological order.
\subsection{On the horizon surface: correlations and bulk entropy}

We start with early results on spin networks describing finite regions of 3D space bounded by a causal horizon. On one hand, these results deal with the computation of the horizon entropy and the recognition of correlations between horizon subregions as responsible for corrections to the entropy area law \cite{Livine:2005mw}; on the other, they concern the introduction of the concept of bulk entropy and its relationship with the boundary area \cite{Livine:2007sy}.\smallskip

\begin{figure}[t]
\centering
\subfigure[\label{fig:splitbound}]{\includegraphics[width=0.4\linewidth]{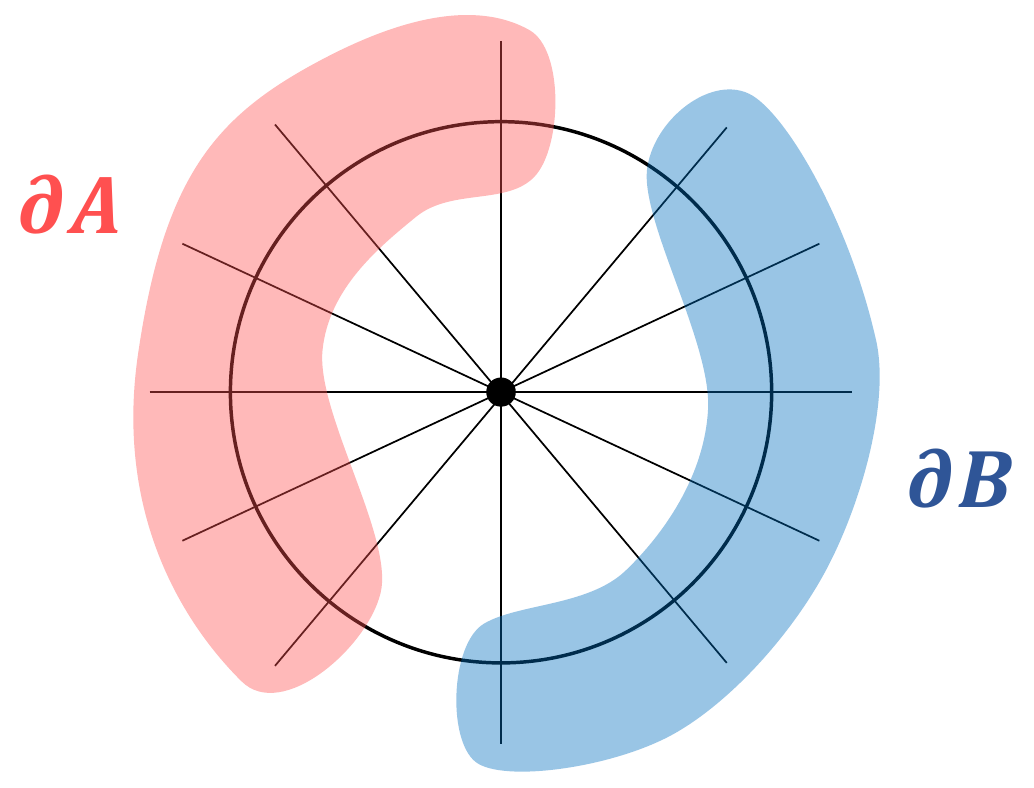}
}
\centering
\subfigure[\label{fig:disj}]{\includegraphics[width=0.52\linewidth]{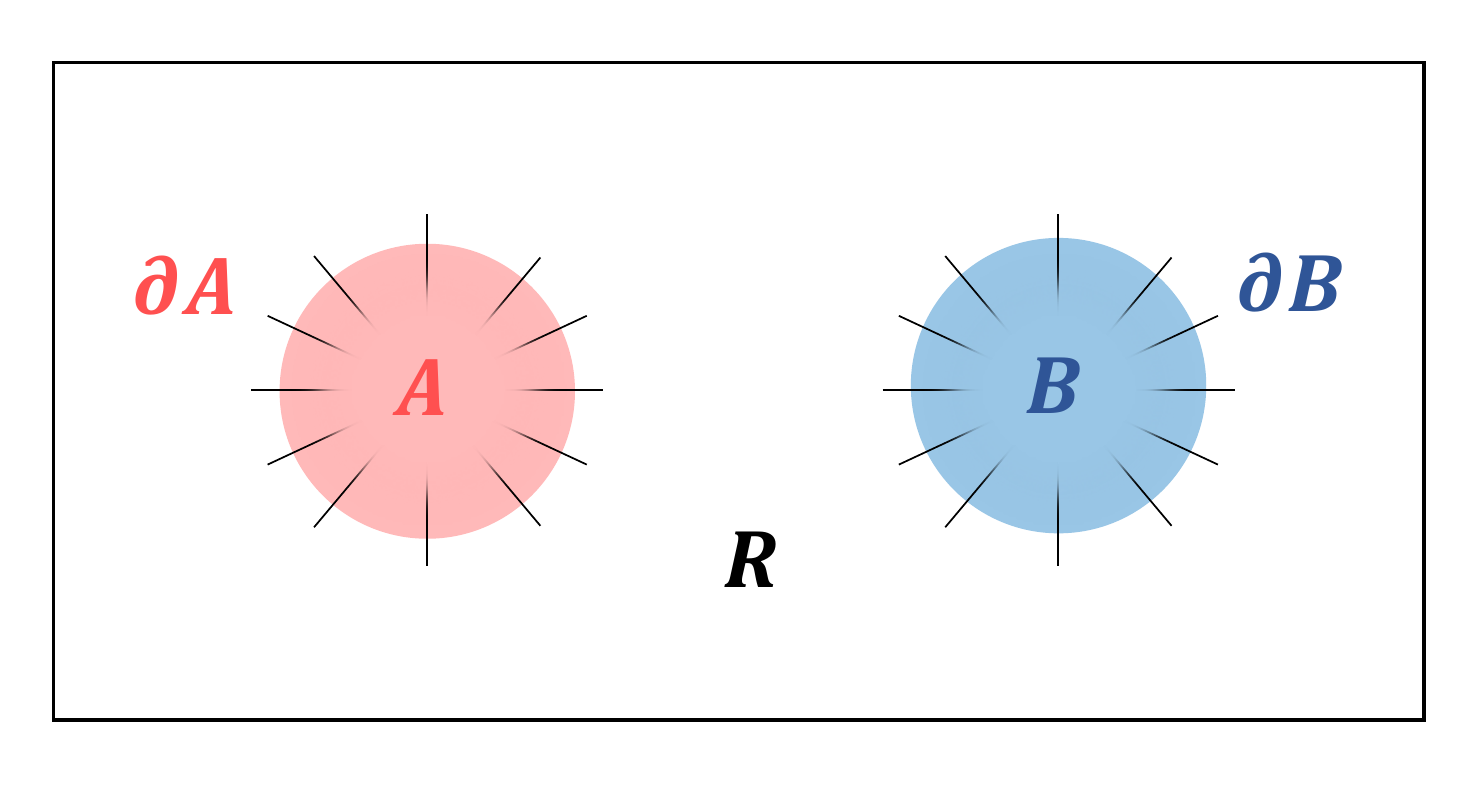}
}
\centering
\subfigure[\label{fig:unfo}]{\includegraphics[width=0.8\linewidth]{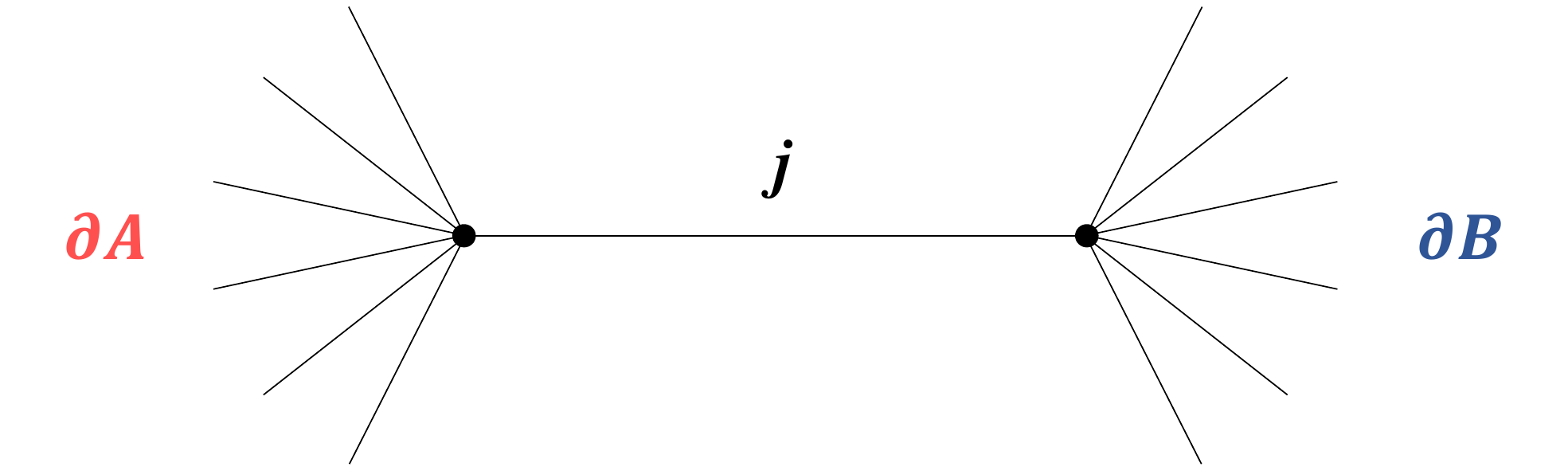}
}
\caption{Studying correlations between parts of the boundary (a) and between disjoint bulk regions (b). Both settings are equivalent to an intertwiner of the spins $\partial A \cup \partial B$, which can be unfolded as depicted in (c): two vertices corresponding to the subsystems $\partial A$ and $\partial B$ are connected by a \virg{fictitious link} representing the correlations between them. For $j=0$ the two subsystems are uncorrelated. 
}
\end{figure}

\paragraph*{Black point model for the computation of the horizon entropy} In Ref.~\onlinecite{Livine:2005mw} Livine and Terno modelled the horizon of a static black hole (at the kinematic level) as a two-sphere made by $2n$ elementary patches, each one punctured by an edge carrying the spin $\frac{1}{2}$ (the argument is as follows: since any representation space $V^j$ can be decomposed into the symmetrised product of $2j$ spin-$\frac{1}{2}$ representations, the spin-$\frac{1}{2}$ patch can be considered as the \virg{elementary patch}). We denote by $R$ the black hole region, so that its boundary $\partial R$ corresponds to the horizon two-sphere. The Hilbert space $\h_{\partial R}$ describing the set of boundary edges can be decomposed as 
\begin{equation}
\h_{\partial R}=\bigotimes^{2n}V^{\frac{1}{2}}\cong\bigoplus_{j=0}^n V^j \otimes D^j_n 
\end{equation}
where $D^j_n$ is the degeneracy space of states with spin $j$. The gauge-invariant subspace associated to the horizon is then given by the intertwiner space
\begin{equation}
\h_{\partial R}^{(0)}=\mathrm{Inv}_{SU(2)}\left[\bigotimes^{2n}V^{\frac{1}{2}}\right] \cong D^0_n
\end{equation}
where the superscript $(0)$ is used to denote the presence of gauge-invariance.
In this description the bulk is thus coarse-grained to a single point (hence the name \virg{black point model}), as depicted in figure \ref{fig:splitbound}. The assumption that the surface is a causal horizon implies complete ignorance of the bulk geometry, and the boundary state is therefore given by 
\begin{equation}
\rho =\frac{1}{N}\sum_r \ket{\iota^r}\bra{\iota^r}
\end{equation}
where $\lbrace\ket{\iota^r}\rbrace$ is a basis of the intertwiner space $\h^0_{\partial R}$ and $N$ the dimension of the latter. Note that the Boltzmann entropy of such state coincides with its von Neumann entropy, both being equal to $\log N$. 
The intertwiner-space dimension $N$ is computed via random walk techniques, and the result for the entropy in the asymptotic limit $n\rightarrow \infty$ is an area law with a logarithmic
correction. The latter is shown to be given by the total amount of correlations between two halves of the horizon surface. Let us show the methodology, as this will be useful for subsequent discussion and for comparing the results presented in part \ref{partIII} to the loop quantum gravity literature.\\
Consider the splitting of the boundary into a set $\partial A$ of $2k$ qubits and a complementary set $\partial B$ of $2(n-k)$ qubits (see figure \ref{fig:splitbound}). Then
\begin{equation}
\h_{\partial R}=\h_{\partial A} \otimes \h_{\partial B}
\end{equation}
where $\h_{\partial A} = (V^{\frac{1}{2}})^{\otimes 2k}$ and $\h_{\partial B}=(V^{\frac{1}{2}})^{\otimes 2(n-k)}$ (note that such a factorisation does not hold for the gauge-invariant subspace $\h_{\partial R}^{(0)}$, see the discussion on Ref.~\onlinecite{Livine:2006xk}). When decomposing each subspace into a direct sum over irreducible representations $j$, e.g. $
\h_{\partial A}=\bigoplus_{j=0} V^j_{\partial A} \otimes D^j_{\partial A}$, the intertwiner states of $\h_{\partial R}^{(0)} \subset \h_{\partial R}$ turn out to be singlet states on $V^j_{\partial A} \otimes V^j_{\partial B}$  with extra indices $a_j$ and $b_j$ labelling basis of the degeneracy spaces $D^j_{\partial A}$ and $D^j_{\partial B}$, respectively. This corresponds to unfolding the intertwiner as illustrated in figure \ref{fig:unfo}. The horizon state then becomes the following:
\begin{equation}
\rho= \frac{1}{N}\sum_j^k \sum_{a_jb_j} \ket{j, a_j, b_j}\bra{j, a_j, b_j}
\end{equation}
It is found that, for $2k=n$ (symmetric splitting of the horizon surface), the mutual information $I_\rho(\partial A:\partial B)$, amounting to three times the entanglement between $\partial A$ and $\partial B$ (quantified e.g.~by the entanglement of formation), equals the logarithmic correction to the horizon entropy.\\
A possible relationship of the entanglement between $\partial A$ and $\partial B$ (for $\partial A \ll \partial B$) with the evaporation process is also suggested, as the case $j=0$ corresponds to the detachment of the surface patch $\partial A$ from the rest of the horizon.\smallskip

\paragraph*{Bulk-topology contribution to the boundary entropy}

In Ref.~\onlinecite{Livine:2007sy} Livine and Terno generalised the computation of the horizon entropy performed in Ref.~\onlinecite{Livine:2005mw} by taking into account the non-trivial structure of the bulk graph. In particular, they promoted the boundary state counting of Ref.~\onlinecite{Livine:2005mw} to a bulk state counting performed by gauge-fixing the holonomies on internal loops to avoid over-estimating the number of states seen by an external observer (it is showed that, because of gauge invariance, the bulk degrees of freedom are truly carried by internal loops). The horizon entropy (evaluated as the logarithm of the number of states supported by a bulk flower-graph with fixed boundary conditions) then turned out to depend on the topology of the graph through its number of loops.\smallskip

\subsection{Distance from entanglement}

\paragraph*{Correlations between disjoint regions of a spin network} In Ref.~\onlinecite{Livine:2006xk} Livine and Terno explored the correlations induced between two disjoint regions $A$ and $B$ of a spin network from the \virg{outside geometry} $R$ (i.e.~the region of the spin network complementary to $A \cup B$, see figure \ref{fig:disj}). Since $\partial A \cup \partial B=\partial R$, the gauge invariant state induced on the boundary of the two regions can be regarded as the result of coarse-graining $R$ to a single intertwiner, as in the model of the previous paragraph. 
The presence of correlations between $A$ and $B$ can then be traced back to the fact that, because of the requirement of gauge invariance of $\partial R$, the Hilbert space $\h^{(0)}_{\partial R}$ is not isomorphic to $\h^{(0)}_{\partial A} \otimes \h^{(0)}_{\partial B}$. The intertwiner on $\partial R$ can in fact be unfolded into two vertices connected by a \virg{fictitious link} as in figure \ref{fig:unfo}, and $\h^{(0)}_{\partial A} \otimes \h^{(0)}_{\partial B}$ is recovered as the subspace with internal link labelled by the trivial representation $j=0$ (which effectively corresponds to the absence of connection). The internal link thus encodes the entanglement between regions $A$ and $B$, induced from the complementary region $R$. This entanglement is then related to a notion of distance between parts of the spin network, building on the idea that, in absence of a background geometry, such a notion can only be defined in term of correlations between the quantum degrees of freedom, and is expected to be induced from the algebraic and combinatorial structure of the \virg{outside geometry}. \\
In the same spirit, Ref.~\onlinecite{Feller:2015yta} by Feller and Livine shows how a notion of distance can be reconstructed from spin network states whose correlations map onto the standard Ising model.\\
Let us finally mention that, as opposed to  Ref.~\onlinecite{Livine:2006xk}, the more recent Ref.~\onlinecite{Livine:2017fgq} by Livine identifies the link entanglement of the unfolded intertwiner as unphysical, as deriving from looking at non-gauge invariant states.

\subsection{Entanglement entropy and holographic spin networks}

Gauge-invariant degrees of freedom are non-local: the Hilbert space of a spin network graph $\h_\gamma$, indeed, does not factorise into the tensor product of Hilbert spaces describing the subgraphs into which $\gamma$ can be split. We mentioned that in Ref.~\onlinecite{Livine:2005mw} and Ref.~\onlinecite{Livine:2006xk} this issue is overcome by embedding the intertwiner space into the tensor product of Hilbert spaces which are \textit{not} gauge-invariant (each one being the tensor product of representations attached to a subset of boundary edges and to a \virg{fictitious} internal-link). Likewise, in Ref.~\onlinecite{Donnelly:2008vx} and Ref.~\onlinecite{Donnelly:2011hn} Donnelly showed how the entanglement entropy between an arbitrary region $R$ of a spin network graph and its complement $\overline{R}$ can be computed by embedding $\h_\gamma$ into an extended Hilbert space that factorises over $R$ and $\overline{R}$, with the gauge symmetry broken at the interface of the two regions. More specifically,  Ref.~\onlinecite{Donnelly:2008vx} takes the complete graph in a spin network basis state: the reduced density matrix $\rho_R$ is therefore completely mixed, and the entropy given by 
\begin{equation}\label{SR}
S(\rho_R)=\sum_{e\in \partial R} (2j_e+1)
\end{equation}
An explanation of the agreement with the result obtained from the isolated horizon framework in the limit of a large number of punctures is then provided: the spin network states representing the purification of $\rho_R$ in the two frameworks have a Schmidt decomposition of the same rank (note that the result holds only asymptotically: the isolated-horizon entropy is less than Eq. \eqref{SR}, as it includes the gauge-invariance constraint on the boundary $\partial R$). Reference \onlinecite{Donnelly:2011hn}, instead, takes the whole graph in a completely generic state. The entropy of region $R$ then turns out to be given by the sum of three positive terms: the Shannon entropy of the distribution of boundary representations, the weighted average of $\log (2j+1)$ over all boundary representations $j$, and a term representing non-local correlations. 

An alternative definition of entanglement entropy of regions of a spin network, similarly derived from the embedding of the Hilbert space of gauge-invariant states into an extended Hilbert space, is provided in Ref.~\onlinecite{Delcamp:2016eya}, and relies on an extension procedure that is based on the excitation content of the theory instead of the underlying graph.

The computation of the entanglement entropy of spin network states and the study of a holographic regime via models and techniques from condensed matter physics is the methodology underlying the results on random spin networks to which this review is dedicated. It has been adopted in earlier work: in Ref.~\onlinecite{Feller:2017jqx} Feller and Livine introduced a class of states inspired by the Kitaev’s toric code model which satisfy an area law for entanglement entropy and whose correlation functions between distant spins are non-trivial. 

We close this subsection with a general result on boundaries in quantum gravity: in Ref.~\onlinecite{Bianchi:2013toa} Bianchi \textit{et al.} showed that boundary states associated to finite portions of spacetime, representing local gravitational processes with certain initial and final data, are mixed, pointing out that such a feature can be regarded as the consequence of tracing over the correlations between the region and its exterior.

\subsection{Gluing adjacent faces with entanglement}

One of the main points of this review concerns the role of entanglement in the connectivity of space. Here we recall recent results by Bianchi and collaborators on entanglement as a tool for gluing (in the sense specified below) elementary portions of space (spin network vertices). Crucially, this will allow us to differentiate between the various notions of gluing of spin network vertices, and to clarify which degrees of freedom are involved in the corresponding entangling procedures. The variety we refer to stems from the distinction between vector geometries, which are defined below, and twisted geometries, of which tensor networks provide a quantum version. In fact, as explained in section \ref{spinnet}, a spin network describes a quantum twisted geometry in which \virg{neighbouring relations} of quantum polyhedra are codified by links: two intertwiners connected by a link $\ell$ represent neighbouring polyhedra, whose adjacent faces have equal area (determined by the spin $j_\ell$) but different shape and/or orientation, in general. Note that the absence of correlation between the polyhedra of a twisted geometry is translated, at the quantum level of the spin network, to the un-entangled nature of the intertwiner degrees of freedom (i.e.~the quantum geometry of neighbouring polyhedra has uncorrelated fluctuations). In a vector geometry the normals to the adjacent faces of neighbouring polyhedra are instead anti-parallel, i.e.~the two faces adhere to each other, despite the possibly different shape. 

In Ref.~\onlinecite{Baytas:2018wjd} it was shown that a quantum version of vector geometries can be obtained from a spin network graph by entangling the intertwiner degrees of freedom. They introduced a class of states, called \textit{Bell-network states}, constructed by creating between intertwiners at nearest-neighbour nodes the analogous of the spin-spin correlations of a Bell singlet states. These correlations ensure that the normals to the adjacent faces of the corresponding quantum polyhedra are always back-to-back, i.e.~that the face planes are parallel. Then, exactly as a Bell singlet state can be understood as a uniform superposition of back-to-back spins over all space directions, a Bell-network state at fixed spins represents a uniform superposition over all vector geometries. In Ref.~\onlinecite{Bianchi:2018fmq} it was further shown that the entanglement entropy of Bell-network states obeys an area law.

\section{Random spin networks and dual statistical models} \label{partII}

So far we introduced the spin network formalism (shared by several approaches to quantum gravity) to describe regions of quantum space(time), and pointed out that entanglement plays a crucial role in this description: it is at the origin of space connectivity. When facing the problem of extracting continuum gravitational physics from such a fundamental description, a crucial issue to be dealt with is the interplay between quantum correlations among the geometric data and global kinematic (and possibly dynamic) geometric features of the spacetime regions considered. Entanglement entropy turned out to be a key tool in this regard\cite{Bianchi:2012ev, Cao:2016mst,VanRaamsdonk:2010pw,Ryu:2006bv,Ryu:2006ef}.

The computation of the entanglement entropy of spin network states can be highly simplified by the use of random tensor network techniques. This clearly requires to restrict the attention to spin network states given by (superpositions of) random tensor networks. We introduce such a class of states in section \ref{randomspinnetworks}, and dedicate section \ref{RTtec} to illustrate how random tensor network techniques can be used to translate their Rényi entropies into partition functions of a classical Ising model. 

\subsection{Random spin networks}\label{randomspinnetworks}

Consider the class of states (introduced in section \ref{TNper}) which are obtainable from the gluing of a set of vertices each one described by a state $f_v$:

\begin{equation}\begin{split}\label{tn}
 |\psi_\gamma \rangle &= \left(\bigotimes_{\ell \in \gamma}\langle \ell| \right) \bigotimes_v |f_v\rangle \\&=\bigoplus_{\vec{j_ \gamma}}\sum_{\vec{n_\gamma}\vec{\iota_ \gamma}}\left((f_1)^{\vec{j}_1}_{\vec{n}_1 \iota_1}... (f_N)^{\vec{j}_N}_{\vec{n}_N \iota_N}\prod_{a^i_{vw}=1} I_{n_v^in_w^i}\right)\ket{\vec{j_\gamma}~\vec{n_\gamma}~\vec{\iota_ \gamma}}
 \end{split}
\end{equation}
where $\vec{j_\gamma}$, $\vec{n_\gamma}$ and $\vec{\iota_ \gamma}$ are, respectively, spins, magnetic indices and intertwiners attached to the graph $\gamma$, and $I$ is the bivalent intertwiner introduced in Eq.~\eqref{bivalent}. A coarse graining of these states is then implemented via uniform randomisation over the geometric data. The randomisation is performed on each vertex separately. This is a necessary requirement for the entropy calculation to be mapped into the evaluation of the free energy of a statistical model (see below). It is also assumed that the spin network states are peaked on specific values $\vec{j_\gamma}$ of the edge spins. This assumption allows us to work in a fixed spin-sector and thus largely simplifies the calculation. The attention is therefore restricted to states of the form
\begin{equation}\begin{split}\label{tnfixed}
(f_1)^{\vec{j}_1}_{\vec{n}_1 \iota_1}... (f_N)^{\vec{j}_N}_{\vec{n}_N \iota_N}\prod_{a^i_{vw}=1} I_{n_v^in_w^i} \coloneqqrev \left(\psi^{\vec{j}_\gamma}_\gamma\right)_{\vec{n_\gamma}\vec{\iota_ \gamma}}
 \end{split}
\end{equation}
where each tensor $(f)^{\vec{j}}_{\vec{n} \iota}$ is picked randomly from its Hilbert space $\h_{\vec{j}}$ (defined in Eq.~\eqref{hj}) according to the uniform probability distribution. In the following we omit from the r.h.s.~of Eq.~\eqref{tnfixed} the explicit reference to the edge spins $\vec{j}_\gamma$; therefore, unless otherwise stated, $\ket{\psi_\gamma}$ refers to the fixed-spin state of Eq.~\eqref{tnfixed}.

\subsection{Rényi entropy from Ising partition function}\label{RTtec}
\begin{figure}[t]
\centering
\subfigure[\label{fig:rho}]{\includegraphics[width=0.48\linewidth]{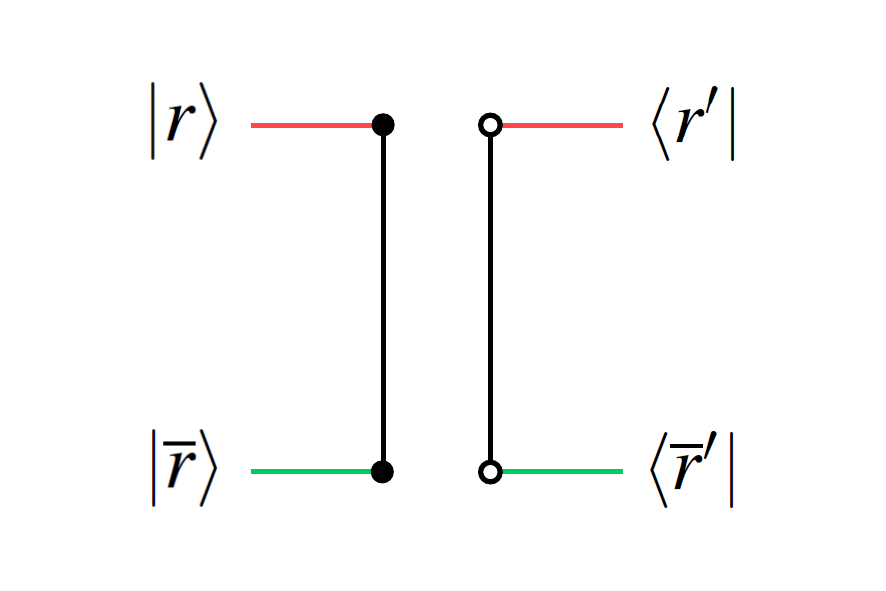}
}
\centering
\subfigure[\label{fig:squared}]{\includegraphics[width=0.49\linewidth]{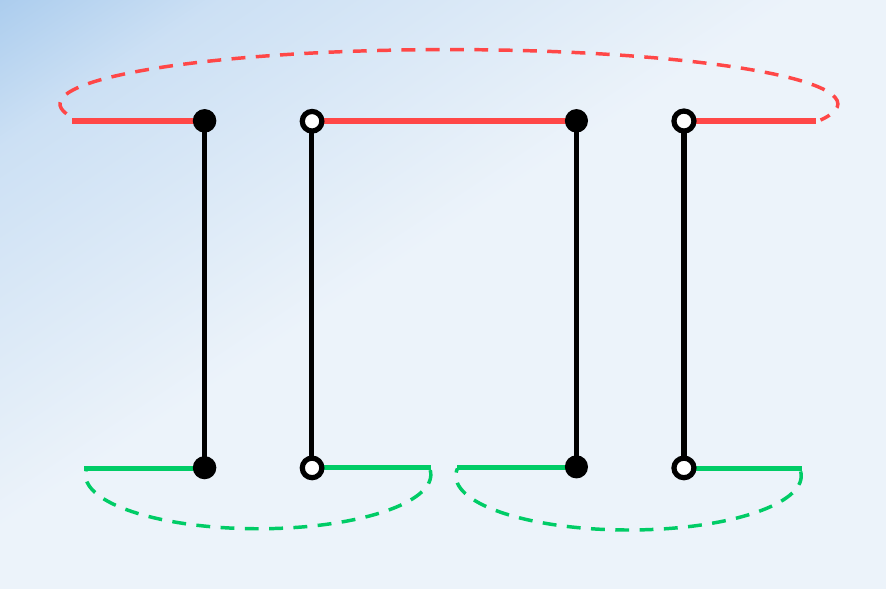}
}
\centering
\subfigure[\label{fig:swap}
]{\includegraphics[width=0.49\linewidth]{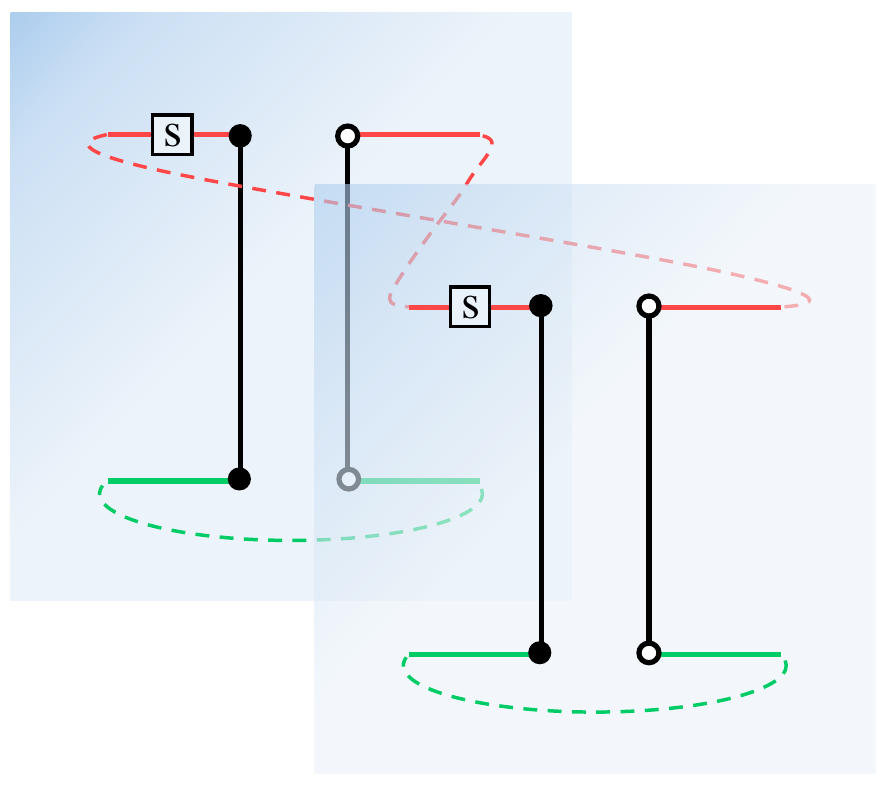}
}
\centering
\subfigure[\label{fig:vertexswap}]{\includegraphics[width=0.48\linewidth]{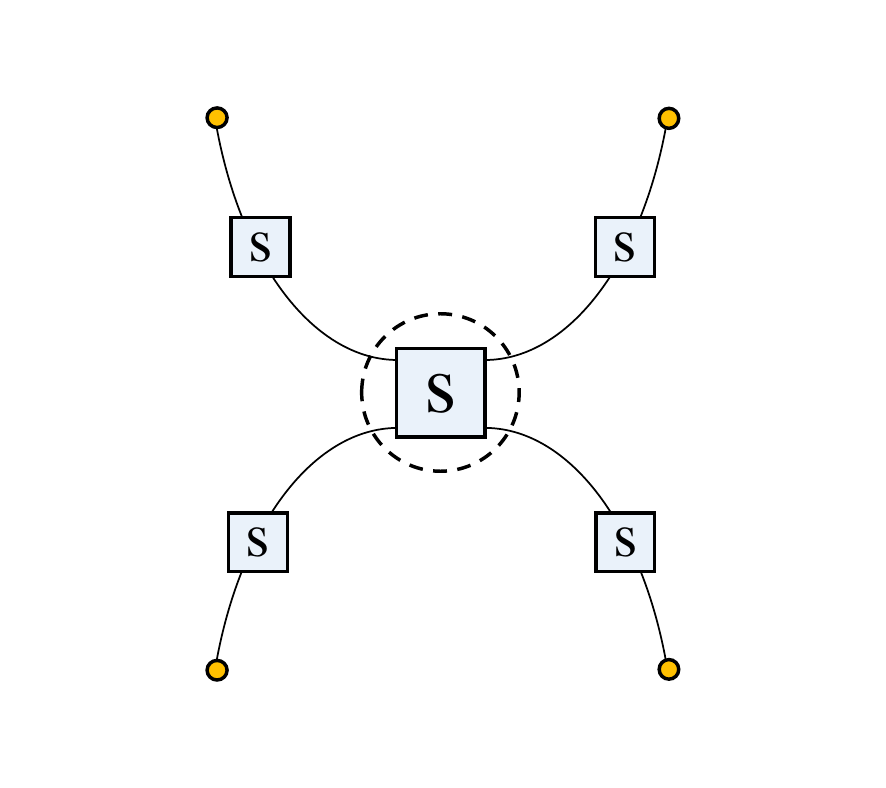}
}
\caption{Illustration of the replica trick in Eq.~\eqref{trrr}. In (a) the state $\rho \in \h_R \otimes \h_{\overline{R}}$: the black disks refer to the subsystems described by $\h_R$ (top) and by $\h_{\overline{R}}$ (bottom), the white disks to the dual components. In (b) the l.h.s. of Eq.~\eqref{trrr}: the trace over $\overline{R}$ (dashed green line) yields $\rho_R$; the latter is then multiplied by itself (connection of internal disks) and traced over (dashed red line). In (c) the r.h.s. of Eq.~\eqref{trrr}: two copies of $\rho_R$ are considered; the swap operator, whose action is denoted by a square, causes the trace to be performed across the two spaces. In (d) the factorization of the swap operator $S$ for the single vertex on the intertwiner (large square in the center) and on each individual edge (small squares)}
\label{freplica}
\end{figure}

As pointed out in section \ref{randomspinnetworks}, our main focus is on finite regions of quantum space described by spin networks corresponding to random tensor networks. In the following we show that the entanglement content of these states can be conveniently computed via the Rényi entropies.

Given the spin network state $\ket{\psi_\gamma}$ of Eq.~\eqref{tnfixed}, consider the reduced state associated to a region $R$ of the graph $\gamma$: $\rho_R=\Tr_{\overline{R}}[\rho]$, where $\rho=\ket{\psi_\gamma}\bra{\psi_\gamma}$ and $\Tr_{\overline{R}}$ is the trace over all degrees of freedom (magnetic indices and/or intertwiners) of the region $\overline{R}$ complementary to $R$. The Rényi-$2$ entropy of $\rho_R$ is a measure of entanglement given by 
\begin{equation}
S_2(\rho_R)\coloneqq\log\Tr(\rho_R^2)
\end{equation}
The computation of this is quantity is performed via the \textit{replica trick}, which is based on the possibility to express the trace of a reduced density matrix $\rho_R$ as a trace over two copies of the density matrix $\rho$ associated to the entire system (here we assumed $\rho$ to be normalised, i.e.~$\Tr(\rho)=1$):
\begin{equation}\label{trrr}
\Tr(\rho_R^2)= \Tr \left[\left(\rho \otimes \rho\right)S_R\right]
\end{equation}
where the operator $S_R$, called \textit{swap operator}, acts on the two copies of the Hilbert space $\h_R$ associated to $R$ as follows:
 \begin{equation}
S_R \ket{r} \otimes \ket{r'}= \ket{r'} \otimes \ket{r}
\end{equation}
 with $\ket{r}$ and $\ket{r'}$ elements of an orthonormal basis of $\h_R$. An illustration of the replica trick of Eq.~\eqref{trrr} is given in figure~\ref{freplica}.

By applying the replica trick, the Rényi-$2$ entropy of the region $R$ of the spin network described by the state $\rho=\ket{\psi_\gamma}\bra{\psi_\gamma}$ can be written as 
\begin{equation}
S_2(\rho_R)=-\log \left(\frac{Z_1}{Z_0}\right), \quad \begin{aligned}
&Z_1 \coloneqq \Tr \left[\left(\rho \otimes \rho\right)S_R\right],\\&Z_0 \coloneqq \Tr \left[\rho \otimes \rho\right],
\end{aligned}
\end{equation}
where the presence of the denominator takes into account the possible non-normalisation of $\rho$, and where the swap operator $S_R$ acts on two copies of the Hilbert space
 \begin{equation}
 \h_R=\left(\bigotimes_{v\in R}\mathcal{I}^{\vec{j}_v}\right)\otimes \left(\bigotimes_{e\in R}V^{j_e}\right),
 \end{equation}
associated to the spin network region $R$.

Note that $Z_1$ and $Z_0$ are quadratic functions of the random vertex states $\rho_v\coloneqq \ket{f_v}\bra{f_v}$, and their average is therefore easier to compute than the average of the entropy. This leads to the proposal\cite{Hayden:2016cfa} of expanding the latter in powers of the fluctuations $\delta Z_1=Z_1 - \overline{Z_1}$ and $\delta Z_0=Z_0 - \overline{Z_0}$ (the overline is used to denote average value under randomisation of the vertex states):
\begin{equation}
\overline{S_2(\rho_R)}= -\log \left(\frac{\overline{Z_1}}{\overline{Z_0}}\right) + \sum_{n=1}^\infty\frac{(-1)^{n-1}}{n}\left(\frac{\overline{\delta Z_0^n}}{\overline{Z_0}^n} - \frac{\overline{\delta Z_1^n}}{\overline{Z_1}^n} \right)
\end{equation}
In Ref.~\onlinecite{Hayden:2016cfa} Hayden \textit{et al.} showed that for large enough bond dimensions, which in the present framework correspond to the edge spins, the fluctuations are suppressed, i.e.~
\begin{equation}\label{aven}
\overline{S_2(\rho_R)}\simeq -\log \left(\frac{\overline{Z_1}}{\overline{Z_0}}\right),
\end{equation}
where $\simeq$ refers to asymptotic equality as the edge spins go to infinity. In particular they proved that, for a tensor network with homogeneous bond dimensions equal to $D$, given an arbitrary small parameter $\delta >0$ it holds 
\begin{equation}
\vert S_2(\rho_R) - \overline{S_2(\rho_R)}\vert < \delta
\end{equation}
with probability $P(\delta)=1-\frac{D_c}{D}$, where $D_c$ is a critical bond dimension depending on $\delta$ and on the number $N$ of vertices as $D_c \propto \delta^{-2}e^{cN}$, with $c$ a constant factor.

Thanks to Eq.~\eqref{aven} the computation of the average entropy can thus be traced back to the computation of the average quantities $\overline{Z_1}$ and $\overline{Z_0}$. Let us focus on $\overline{Z_{1}}$, as $\overline{Z_{0}}$ is simply given by the latter upon reducing the swap operator to the identity operator. The quantity $\overline{Z_{1}}$ can be written as\cite{Colafranceschi:2021acz}
 \begin{equation}\begin{split}\label{z1}
\overline{Z_{1}}= \mathrm{Tr} \left[\left(\bigotimes_\ell\rho_{\ell}^{\otimes 2}\right)\left(\bigotimes_v \overline{\rho_{v}^{\otimes 2}}\right)S_R\right],
 \end{split}
\end{equation}
where $\rho_{\ell}\coloneqq \ket{\ell}\bra{\ell}$ and $\rho_v\coloneqq \ket{f_v}\bra{f_v}$. For each vertex $v$, the average over the two copies of the state $\rho_{v}$ can be computed via the Schur's lemma\cite{2013arXiv1308.6595H}, which yields
\begin{equation}\label{schur}
\overline{\rho_{v}^{\otimes 2}}=\frac{ \mathbb{I} + S_v}{\mathcal{D}_v(\mathcal{D}_v+1)},
\end{equation}
where $\mathcal{D}_v$ is the dimension of the vertex Hilbert space $\h_{\vec{j_v}}$ (see Eq.~\eqref{hj}) and $S_v$ is the swap operator on $\h_{\vec{j_v}}\otimes \h_{\vec{j_v}}$. This is a crucial step since it brings out, as we are going to explain, two-level variables (one for each vertex) corresponding to the spins of a Ising model living on the graph $\gamma$. Note in fact that, once Eq.~\eqref{schur} is inserted into Eq.~\eqref{z1}, the latter can be written as 
\begin{equation}\begin{split}\label{zisigma}
\overline{Z_{1}}&=\mathcal{C} ~\sum_{\vec{\sigma}}\mathrm{Tr} \left[\left(\bigotimes_\ell\rho_{\ell}^{\otimes 2}\right)\left(\bigotimes_{v} S_v^{\frac{1-\sigma_v}{2}}\right)S_R\right],
 \end{split}
\end{equation}
where $\sigma_v=\pm 1$ is a two-level variable associated to vertex $v$, $\vec{\sigma}=\{\sigma_1,...,\sigma_N\}$ and
\begin{equation}
\mathcal{C}\coloneqq\prod_v\frac{1}{\mathcal{D}_v(\mathcal{D}_v+1)}
\end{equation}
is a constant factor. That is, $\overline{Z_{1}}$ has been written as a sum of $2^N$ terms involving the identity ($\mathbb{I}$) or the swap operator ($S_v$) for each of the $N$ vertices, and the variable $\sigma_v$ encodes the presence of one or the other ($\mathbb{I}$ for $\sigma_v=+ 1$ and $S_v$ for $\sigma_v=- 1$) in every term of the sum.

Given the form of the vertex Hilbert space $\h_{\vec{j_v}}$, the swap operator $S_v$ factorises as follows:
\begin{equation}\label{swapfactorized}
S_v=\bigotimes_{i=0}^d S_v^i,
\end{equation}
i.e.~into a swap operator $S_v^0$ for (the double copy of) the intertwiner Hilbert space $\mathcal{I}^{\vec{j}_v}$ and a swap operator $S_v^i$ for (the double copy of) the representation space $V^{j_v^i}$ on each edge $e_v^i$, as shown in figure \ref{fig:vertexswap}. Crucially, the same applies to the swap operator $S_R$:
\begin{equation}\begin{split}
S_R=\left(\bigotimes_{e_v^i\in R}  S_v^i\right)\left(\bigotimes_{v\in R}  S_v^0\right).
 \end{split}
\end{equation}
Consequently, to every open edge $e_v^i$ of the graph $\gamma$ one can attach a two-level variable $\mu_v^i=\pm 1$ (also called \textit{pinning spin}\cite{Hayden:2016cfa}) encoding whether ($\mu_v^i=-1$) or not ($\mu_v^i=+1$) an additional swap operator acts on (the double copy of) its Hibert space; that is, whether or not it belongs to region $R$. The same holds true for the intertwiner on each vertex $v$ of the graph, for which the two-level variable $\nu_v=\pm 1$ is introduced.

By performing the trace in Eq.\eqref{zisigma} one finally obtains that the quantity $\overline{Z_{1}}$ corresponds to the partition function of a classical Ising model:
\begin{equation}\begin{split}
\overline{Z_{1}}&=\sum_{\vec{\sigma}}e^{-A_{1}\left(\vec{\sigma}\right)}
 \end{split}
\end{equation}
with $A_{1}(\vec{\sigma})$ the Ising action
\begin{multline}\label{A1}
    A_{1}\left(\vec{\sigma}\right)= \sum_{\ell_{vw}^i\in \gamma}\frac{1-\sigma_v\sigma_w}{2}\log d_{j_{vw}^i}
+\sum_{e_v^i\in \partial \gamma}\frac{1-\sigma_v\mu_v^i}{2}\log d_{j_v^i}\\+\sum_v\frac{1-\sigma_v\nu_v}{2}\log D_{\vec{j}_v} + const ~,
\end{multline}
where $d_j$ is the dimension of the representation space $V^j$, and $D_{\vec{j}}$ the dimension of the intertwiner space $\mathcal{I}^{\vec{j}}$ (see section \ref{quantumofspace}). Note that the Ising model is defined on the graph $\gamma$:  Eq.~\eqref{A1} involves interactions between nearest neighbours Ising spins, where the adjacency relationship is determined by $\gamma$ (two Ising spins interact only if the corresponding vertices are connected by a link); every Ising spin also interacts with the pinning spins located at its vertex (e.g. the Ising spin $\sigma_v$ of a vertex $v$ on the boundary interacts with the pinning field $\nu_v$ on the intertwiner of $v$ and with the pinning field $\mu_v^i$ on the open edge $e_v^i$ of $v$). 

As far as $\overline{Z_{0}}$ is concerned, we pointed out that it corresponds to $\overline{Z_{1}}$ with $R=\emptyset$ (in fact $S_\emptyset = \mathbb{I}$). Therefore it holds that $\overline{Z_{0}}=\sum_{\vec{\sigma}}e^{-A_{0}\left(\vec{\sigma}\right)}$, where $A_{0}$ is given by Eq.~\eqref{A1} with all pinning spins equal to $+1$:
\begin{multline}\label{A0}
    A_{0}\left(\vec{\sigma}\right)= \sum_{\ell_{vw}^i\in \gamma}\frac{1-\sigma_v\sigma_w}{2}\log d_{j_{vw}^i}
+\sum_{e_v^i\in \partial \gamma}\frac{1-\sigma_v}{2}\log d_{j_v^i}\\+\sum_v\frac{1-\sigma_v}{2}\log D_{\vec{j}_v} + const~.
\end{multline}
Note also that, since $\overline{Z_{0}}$ and $\overline{Z_{1}}$ enter $\overline{S_2(\rho_R)}$ only via their ratio, in the computation of the entropy the constant factor in Eq.~\eqref{A1} and Eq.~\eqref{A0} is irrelevant; we therefore omit it in the following.


To study the properties of the partition function $\overline{Z_{1}}$ it is useful to rewrite the Ising action $A_1(\vec{\sigma})$ in the form $A_{1}(\vec{\sigma})=\beta   H_{1}(\vec{\sigma})$, where $\beta\coloneqq d_j$ with $j$ the average spin on $\gamma$, and
\begin{equation}\begin{split}\label{H1}
  H_{1}\left(\vec{\sigma}\right)=& \sum_{\ell_{vw}^i\in \gamma}\frac{1-\sigma_v\sigma_w}{2}\frac{\log d_{j_{vw}^i}}{\beta}
+\sum_{e_v^i\in \partial \gamma}\frac{1-\sigma_v\mu_v^i}{2}\frac{\log d_{j_v^i}}{\beta}\\&+\sum_v\frac{1-\sigma_v\nu_v}{2}\frac{\log D_{\vec{j}_v}}{\beta} .
\end{split}
\end{equation}
The parameter $\beta$ then plays the role of inverse temperature of the Ising model. As we are working in the high spins regime, the partition function $\overline{Z_{1}}$ is dominated by the lowest energy configuration:
\begin{equation}
\overline{Z_{1}}\simeq e^{-\beta \min_{\vec{\sigma}}H_{1}\left(\vec{\sigma}\right)}.
\end{equation}
The same applies to $\overline{Z_{0}}$ and, since $\min_{\vec{\sigma}}H_{0}=0$ (where $H_0$ is given by Eq.~\eqref{H1} with $\mu_v^i=\nu_v=+1$ $\forall v, e_v^i \in \gamma$), it holds that
\begin{equation}
\overline{Z_{0}}\simeq e^{-\beta \min_{\vec{\sigma}}H_{0}\left(\vec{\sigma}\right)}=1.
\end{equation}
Therefore, the average entropy can be finally computed via the following formula:
\begin{equation}
\overline{S_2(\rho_R)}\simeq -\log \left(\frac{\overline{Z_1}}{\overline{Z_0}}\right)\simeq\beta \min_{\vec{\sigma}}H_{1}\left(\vec{\sigma}\right),
\end{equation}
with $\beta$ the average dimension of the edge spins and $H_1(\vec{\sigma})$ the Ising-like Hamiltonian defined in Eq.~\eqref{H1}.
\section{Holographic entanglement in spin network states}\label{partIII}

We present recent works that explored the connection between holographic features of regions of quantum space and entanglement of their quantum geometric data, for spin network states obtainable from the gluing of random vertex states.
\subsection{Bulk-to-boundary quantum channels: isometric mapping of quantum-geometric data}\label{btob}

Reference~\onlinecite{Colafranceschi:2021acz} analysed the flow of information from the bulk to the boundary of regions of quantum space described by the class of spin network states defined in Eq.~\eqref{tnfixed}, to determine under which conditions such a flow can be holographic. 

Let us start by providing the definitions of bulk and boundary of a spin network, as given in Ref.~\onlinecite{Colafranceschi:2021acz}. 
Consider a spin network with combinatorial pattern $\gamma$ and edge spins $\vec{j_\gamma}$. The \textit{boundary} consists in the set of open edges of $\gamma$ (denoted by $\partial \gamma$) decorated by the respective spins, and is described by the Hilbert space 
\begin{equation}
\h_{\partial \gamma}\coloneqq \bigotimes_{e \in \partial \gamma} V^{j_e};
\end{equation}
let $\ket{\underline{n}}\coloneqq \bigotimes_{e \in \partial \gamma} \ket{j_e n_e}$ be the basis element of the boundary space $\h_{\partial \gamma}$.
The \textit{bulk} is the set of vertices of $\gamma$ (denoted by $\dot{\gamma}$) together with the intertwiners attached to them, and is described by the Hilbert space
\begin{equation}
\h_{\dot{\gamma}}\coloneqq\bigotimes_v \mathcal{I}^{\vec{j}_v};
\end{equation}
let $\ket{\underline{\iota}}\coloneqq \bigotimes_v \ket{\vec{j}_v \iota_v}$ be the basis element of the bulk space $\h_{\dot{\gamma}}$.

The flow of information from the bulk to the boundary is identified with the bulk-to-boundary map that every spin network state implicitly defines once regarding the bulk space as \textit{input} and the boundary space as \textit{output}. More specifically, every spin network state of the form
\begin{equation}
\ket{\phi_\gamma}=\sum_{\underline{n}\underline{\iota}} \left(\phi_\gamma\right)_{\underline{n}\underline{\iota}} \ket{\underline{n}}\ket{\underline{\iota}},
\end{equation}
(to simplify the notation, we omitted the edge spins, as they are fixed) can be regarded as a \textit{map} $\mathcal{M}$ from the bulk to the boundary Hilbert space, having components 
\begin{equation}
\bra{\underline{n}}\mathcal{M}\ket{\underline{\iota}}=\left(\phi_\gamma\right)_{\underline{n}\underline{\iota}}.
\end{equation}
The map $\mathcal{M}$ associated to $\ket{\phi_\gamma}$ therefore acts on a generic bulk state $\ket{\zeta}\in \h_{\dot{\gamma}}$ as follows: 
\begin{equation}
\mathcal{M} \ket{\phi_\gamma} = \bra{\zeta} \phi_\gamma \rangle
\end{equation}
i.e.~by evaluating the spin network state on $\ket{\zeta}$ or, in tensor network language, by feeding the bulk input with $\ket{\zeta}$ (see figure \ref{tensore_fig}).

\begin{figure}[t]
	\centering
	\includegraphics[width=0.8\linewidth]{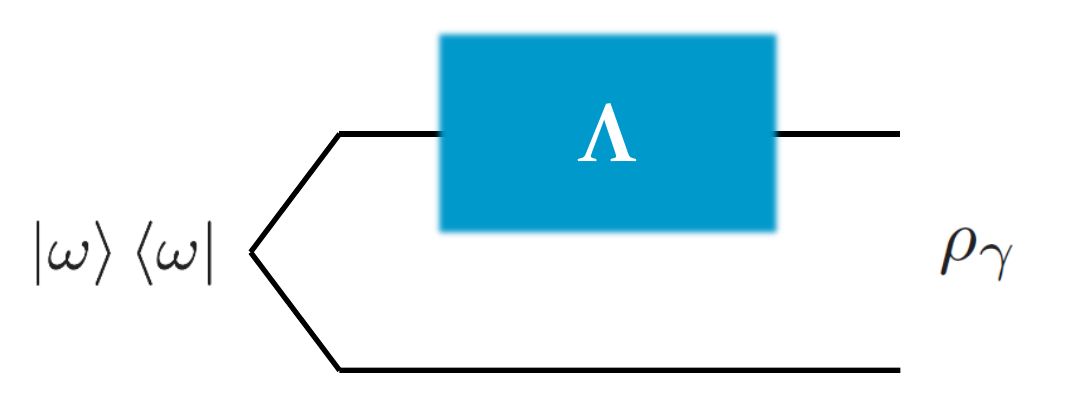}\caption{Relationship between a spin network state $\rho_\gamma$ and the corresponding bulk-to-boundary superoperator $\Lambda$; $\ket{\omega}$ is a maximally entangled state of two bulk copies. }
	\label{channel_fig}
\end{figure}

The reduced (and normalised) bulk state takes the form
\begin{equation}\begin{split}\label{reduced}
  \rho_{\dot{\gamma}}&\coloneqq\frac{1}{D_{\dot{\gamma}}}\text{Tr}_{\partial \gamma} \left[\rho_\gamma\right]\\&=\frac{1}{D_{\dot{\gamma}}} \sum_{\underline{\iota}\underline{\iota'}} \left(\mathcal{M}^\dagger  \mathcal{M} \right)_{\underline{\iota'}\underline{\iota}}|\underline{\iota}\rangle\langle\underline{\iota'}|
\end{split}
\end{equation}
where $\rho_\gamma=\ket{\phi_\gamma}\bra{\phi_\gamma}$ and $D_{\dot{\gamma}}$ is the dimension of the bulk Hilbert space $\h_{\dot{\gamma}}$. It follows from Eq.~\eqref{reduced} that if the reduced bulk state is maximally mixed, namely $\rho_{\dot{\gamma}}=\frac{\mathbb{I}}{D_{\dot{\gamma}}}$, the map $\mathcal{M}$ is an isometry, i.e.~$\mathcal{M}^\dagger  \mathcal{M}=\mathbb{I}$. Moreover, the corresponding superoperator on the space of bulk operators, $\Lambda(\cdot)\coloneqq\mathcal{M} \cdot \mathcal{M}^\dagger$, is a completely positive trace preserving (CPTP) map, with Choi-Jamio\l kowski state 
\begin{equation}\begin{split}
J(\Lambda)=\Lambda \otimes \mathbb{I} \left(  \frac{\ket{\omega}\bra{\omega}}{D_{\dot{\gamma}}}\right)=\frac{\rho_\gamma}{D_{\dot{\gamma}}}
\end{split}
\end{equation}
where
\begin{equation}\begin{split}
|\omega\rangle= \sum_{\underline{\iota}}|\underline{\iota}\rangle \otimes  |\underline{\iota} \rangle
\end{split}
\end{equation}
is a maximally entangled state of two copies of the bulk (see figure \ref{channel_fig}).

Reference~\onlinecite{Colafranceschi:2021acz} studied the bulk-to-boundary map $\mathcal{M}$ of a spin network state of the form of Eq.~\eqref{tnfixed}, to analyse the relationship between the combinatorial structure and geometric data of a spin network on the one hand, and the isometric character of the corresponding map on the other. The latter is quantified via the Rényi-2 entropy of the reduced bulk state (see Eq.~\eqref{reduced}). Thanks to the random nature of the vertex tensors, the entropy is computed via an Ising partition function, according to the technique illustrated in section \ref{RTtec}. In particular, 
\begin{equation}
\overline{S_2(\rho_{\dot{\gamma}})}= \beta \min_{\vec{\sigma}}H_{1}\left(\vec{\sigma}\right)
\end{equation}
with $H_{1}\left(\vec{\sigma}\right)$ the Ising-like Hamiltonian
\begin{equation}\begin{split}\label{H1map}
H_{1}\left(\vec{\sigma}\right)=& \sum_{\ell_{vw}^i\in \gamma}\frac{1-\sigma_v\sigma_w}{2}\frac{\log d_{j_{vw}^i}}{\beta}
+\sum_{e_v^i\in \partial \gamma}\frac{1-\sigma_v}{2}\frac{\log d_{j_v^i}}{\beta}\\&+\sum_v\frac{1+\sigma_v}{2}\frac{\log D_{\vec{j}_v}}{\beta} .
\end{split}
\end{equation}
It is found that spin network graphs made of four-valent vertices (dual to 3D spatial geometries) with an homogeneous assignment of edge spins does not realise an isometric mapping of data from the bulk to boundary. Coherently, increasing the inhomogeneity of the spins assigned to a spin network with four-valent vertices increases the \virg{isometry degree} of the corresponding bulk-to-boundary map.

Let us close this section by commenting on the comparison of this work with Ref.~\onlinecite{Chen:2021vrc}, where the idea of interpreting spin network states as maps from the bulk to the boundary first appeared. In Ref.~\onlinecite{Chen:2021vrc}, Chen and Livine pointed out that spin network wavefunctions with support on an open graph can be regarded as linear forms on the boundary Hilbert space (the space of spin states living on the open edges of the spin network), and that coarse-graining the bulk, i.e.~integrating over the bulk holonomies, then induces a probability distribution for the boundary degrees of freedom. Based on that, they proved the following:  any boundary density matrix can be obtained, via the bulk-to-boundary coarse-graining procedure, from a pure bulk state with support on a graph composed of a single vertex connecting all boundary edges to a single bulk loop. A crucial difference between the map of Ref.~\onlinecite{Chen:2021vrc} and $\mathcal{M}$ is that the latter does not perform a coarse graining of the bulk (intended as tracing out the bulk holonomies); instead, it \textit{evaluates} the (pure) spin network state on a given bulk configuration (specifically, a given state for the intertwiner degrees of freedom), thereby yielding a boundary state. Consequently, the latter is a pure state if the bulk input state is pure. By contrast, the boundary density matrix resulting from the bulk-to-boundary coarse-graining of Ref.~\onlinecite{Chen:2021vrc} applied to a pure spin network state is typically mixed.

\subsection{Holographic states and black hole modelling}\label{bstates}
\begin{figure}[t]
	\centering
	\includegraphics[width=0.4\linewidth]{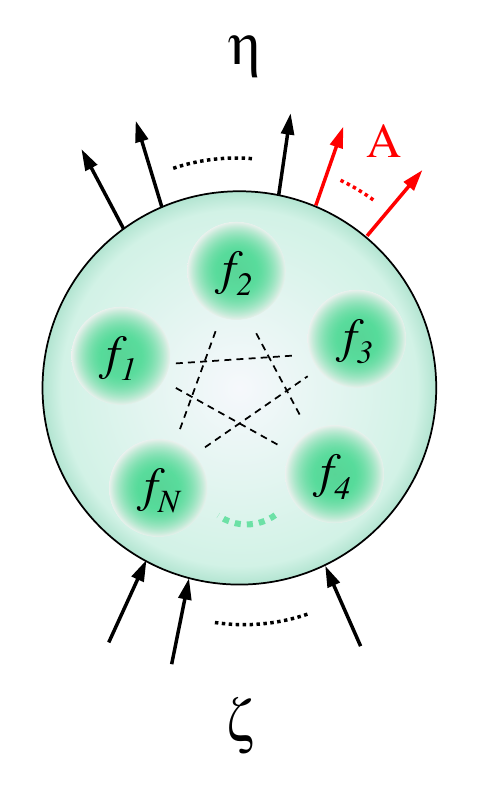}\caption{Spin network state given by the gluing (symbolised by the dotted lines) of random vertex tensors $f\in\h_{\vec{j}}$ (the green disks). $\zeta$ is the input state for the bulk degrees of freedom (intertwiners), graphically depicted as black input lines; $\eta$ is the output state for the boundary edges, depicted as output lines. The boundary entanglement entropy is computed for a set $A$ of the latter, shown in red.} 
	\label{tensore_fig}
\end{figure}

\begin{figure}[t]
	\begin{minipage}[t]{\linewidth}
		\centering
		\includegraphics[width=0.5\linewidth]{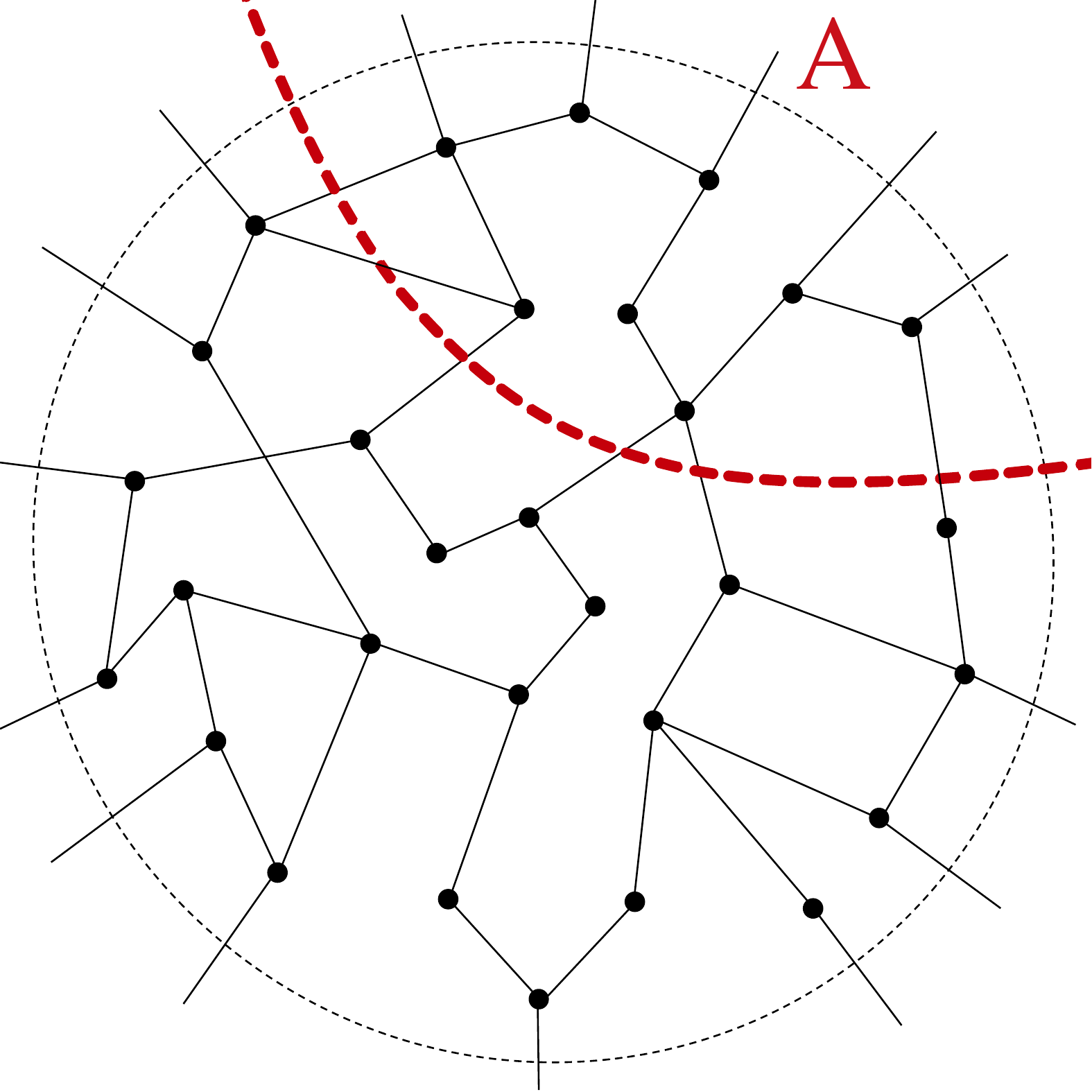}\caption{Area law for the Rényi-2 entropy of a portion $A$ of the boundary of the spin network state in Eq.~\eqref{eta}. The dotted red line represents the Ising domain wall $\Sigma(\vec{\sigma})$.\vspace{0.5cm}}
		\label{inn}
	\end{minipage}
	
	\begin{minipage}[t]{\linewidth}
		\centering
		\includegraphics[width=0.5\linewidth]{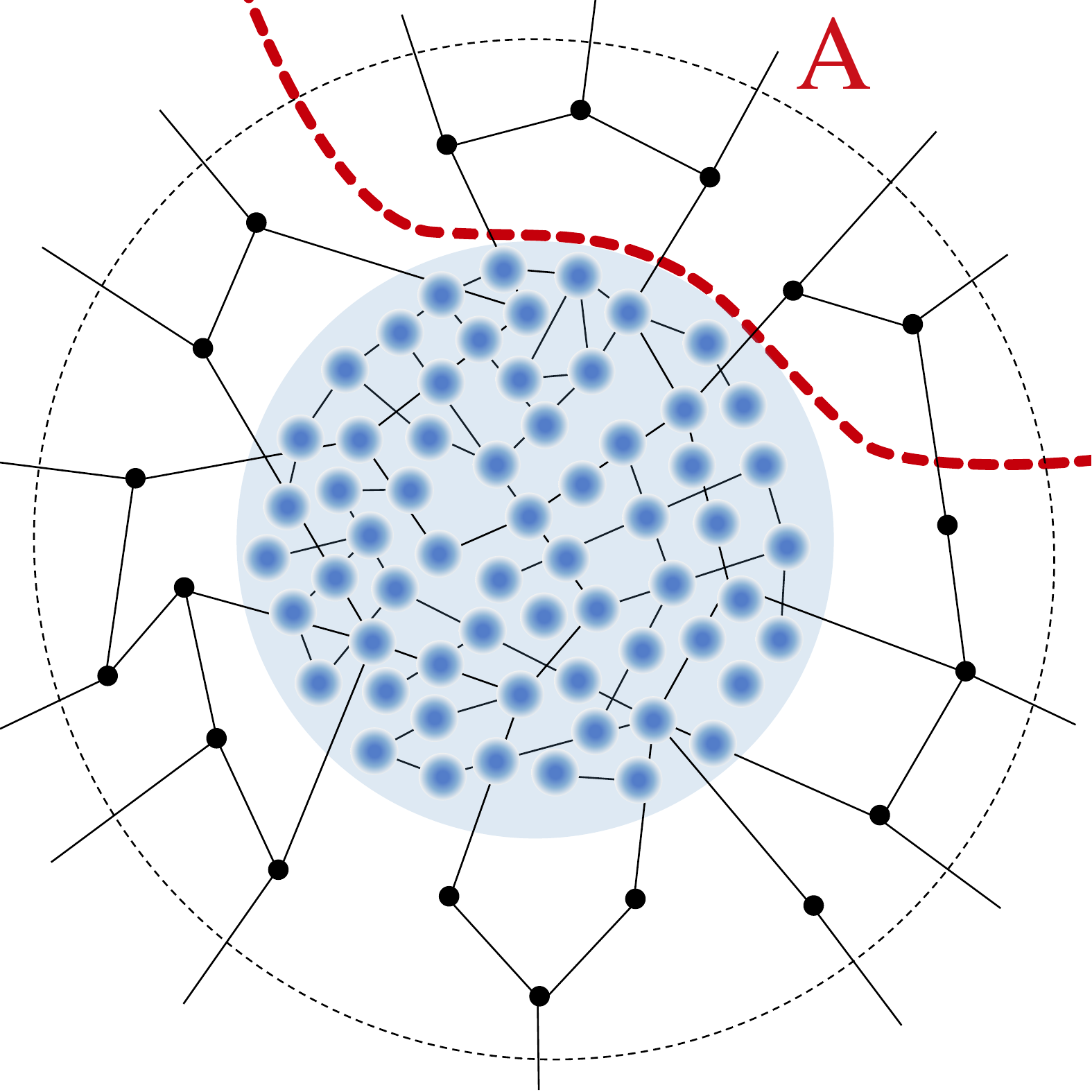}\caption{Emergence of a horizon-like surface in the bulk: when the entanglement entropy of the intertwiners in a region of the graph (the blue disk) exceeds a certain threshold, that region becomes inaccessible to the Ising domain wall $\Sigma(\vec{\sigma})$ (represented by the dotted red line).}
		\label{outt}
	\end{minipage}
\end{figure}

As illustrated in section \ref{btob}, Ref.~\onlinecite{Colafranceschi:2021acz} investigated holography on spin network states having the form of Eq.~\eqref{tnfixed}, regarding them as maps from the bulk to the boundary. Inspired by similar questions, Ref.~\onlinecite{Chirco:2021chk} studied the same class of states from a different perspective: it analysed the boundary states returned by the bulk-to-boundary map, on varying the bulk input state. That is, 
\begin{equation}\begin{split}\label{eta}
\ket{\eta} &= \mathcal{M}\ket{\zeta} \\&=\bra{\zeta} \psi_\gamma \rangle ,
\end{split}
\end{equation}
where $\mathcal{M}$ is the bulk-to-boundary map corresponding to the spin network state (and random tensor network) $\ket{\psi_\gamma}$, defined in Eq.~\eqref{tnfixed}; $\ket{\zeta}\in \h_{\dot{\gamma}}$ is the input bulk state and $\ket{\eta}\in \h_{\partial\gamma}$ the output boundary state. In particular, it focused on the entanglement content of a portion $A$ of the output boundary state (see figure~\ref{tensore_fig}). 
Again, given the random character of the state, the entanglement measure considered is the Rényi-2 entropy, computed via the Ising model. The result is the following:
\begin{equation}\label{etaA}
\overline{S_2(\eta_A)}= \beta \min_{\vec{\sigma}}H_{1}\left(\vec{\sigma}\right)
\end{equation}
where
\begin{equation}\begin{split}\label{H1boundaryA}
H_{1}\left(\vec{\sigma}\right)=& \sum_{\ell_{vw}^i\in \gamma}\frac{1-\sigma_v\sigma_w}{2}\frac{\log d_{j_{vw}^i}}{\beta}
+\sum_{e_v^i\in \partial \gamma}\frac{1-\sigma_v\mu_v^i}{2}\frac{\log d_{j_v^i}}{\beta} \\&+ \frac{1}{\beta}S_2(\zeta_\downarrow)
\end{split}
\end{equation}
with $\zeta_\downarrow$ the bulk state reduced to the region with Ising spins pointed down. From Eq.~\eqref{H1boundaryA} one can note that every misalignment between the Ising spins $\sigma_v$ and $\sigma_w$ on a link $\ell_{vw}^i$ carries a contribution to the entropy equal to $\big({\log d_{j_{vw}^i}}\big)/\beta$, i.e.~to (the logarithm of) the dimension of that link, normalised by $\beta$ (the average value that  quantity can take). The same holds for the pinning spin $\mu_v^i$ and the Ising spin $\sigma_v$ on a boundary edge $e_v^i$. As a result, the first two terms of the r.h.s.~of Eq.~\eqref{H1boundaryA} provide the \virg{area} of the Ising domain wall, i.e.~the surface separating the spin-down region (externally bounded by $A$) from the spin-up region, where the area is given by a weighted sum of the links crossing it (with weights proportional to the logarithm of the link dimensions). Let $\Sigma\left(\vec{\sigma}\right)$ be the aforementioned surface for the Ising configuration $\vec{\sigma}$, and
\begin{equation}
|\Sigma\left(\vec{\sigma}\right)|\coloneqq  \sum_{\ell_{vw}^i\in \gamma}\frac{1-\sigma_v\sigma_w}{2}\frac{\log d_{j_{vw}^i}}{\beta}
+\sum_{e_v^i\in \partial \gamma}\frac{1-\sigma_v\mu_v^i}{2}\frac{\log d_{j_v^i}}{\beta} 
\end{equation}
its area, as defined above. The Ising Hamiltonian of Eq.~\eqref{H1boundaryA} can then be written as follows:
\begin{equation}\begin{split}\label{H1Sigma}
H_{1}\left(\vec{\sigma}\right)=|\Sigma\left(\vec{\sigma}\right)|+ \frac{1}{\beta}S_2(\zeta_\downarrow).
\end{split}
\end{equation}
Combining Eq.~\eqref{etaA} with Eq.~\eqref{H1Sigma} one then finds that, for $S_2(\zeta_\downarrow)\ll \beta \Sigma(\vec{\sigma})$, the Rényi-2 entropy follows an area law with a small correction deriving from the bulk entanglement (see figure \ref{inn}):
	\begin{equation}
	\overline{S_2(\eta_A)}=\beta \left(\min_{\vec{\sigma}} |\Sigma(\vec{\sigma})| \right)+ S_2(\zeta_\downarrow).
	\end{equation}
	For $S_2(\zeta_\downarrow)=O\left( \beta \Sigma(\vec{\sigma})\right)$, instead, the Rényi-2 entropy follows an \virg{area$+$volume law}:
	\begin{equation}
	\overline{S_2(\eta_A)}=\beta \min_{\vec{\sigma}}\lbrace |\Sigma(\vec{\sigma})| +\frac{1}{\beta} S_2(\zeta_\downarrow)\rbrace .
	\end{equation}
In fact, $\overline{S_2(\eta_A)}$ depends to a comparable extent on the entanglement content of the surface $\Sigma(\vec{\sigma)}$ (link entanglement) and of the spin-down region bounded by it (intertwiner entanglement in $\zeta$). 
	
In Ref.~\onlinecite{Chirco:2021chk} it was also showed that increasing the entanglement content of a region of the bulk can turn the boundary of that region into a horizon-like surface (see figure \ref{outt}), as the Ising domain wall which determines the entropy cannot access it. Notably, this result can be regarded as a realisation of the proposal made by Krasnov and Rovelli in Ref.~\onlinecite{Krasnov:2009pd} of defining a quantum black hole as the part of a spin network that does not influence observables at infinity.

\section{Discussion}

We reviewed very recent work that contributes to the research effort which looks at holography not as an asymptotic global property (as it was originally conceived), but as a quasi-local property entering the description of finite spatial boundaries (spacetime corners). The main novelty of the illustrated approach is the use of spin network states formally corresponding to (generalised) random tensor networks. The defining feature of this class of states is the randomness of the wavefunctions associated to the individual spin network vertices, which has the remarkable property of mapping the correlations of the spin network states to that of a classical Ising model living on the same graph. This enables to investigate the entanglement content of the spin network by relying on standard condensed matter and quantum information techniques. Moreover, the randomization over vertex wavefunctions can be understood as a local coarse graining on the vertex data and thus makes this type of states of immediate interest for GFT cosmology\cite{ Gielen:2013kla,Gielen:2013naa,Gielen:2016dss,Oriti:2015qva}.

Reference~\onlinecite{Colafranceschi:2021acz} specifically studied the flow of information from the bulk to the boundary through the Choi-Jamio\l kowski duality, computing the Rényi entropy of the Choi-Jamio\l kowski state through a random tensor technique that traces it back to the evaluation of Ising partition functions. The result is a positive correlation between the inhomogeneity of the edge spins and the \virg{isometry degree} of the bulk-to-boundary map. The same technique is applied in Ref.~\onlinecite{Chirco:2021chk} to the computation of the Rényi entropy of boundary states, and leads to the derivation of (an analogue of) the Ryu-Takayanagi formula\cite{Ryu:2006bv,Ryu:2006ef}. Interestingly, Ref.~\onlinecite{Chirco:2021chk} also showed that the presence of a bulk region with high entanglement entropy can turn the boundary of that region into a horizon-like surface, hereby offering a concrete example of the definition of quantum black holes given in Ref.~\onlinecite{Krasnov:2009pd},  with a picture that recalls the \virg{quantum graphity} of Ref.~\onlinecite{Konopka:2008hp}.

The illustrated work paves the way to an extensive application of quantum information tools to the study of the spacetime microstructure and the modelling of quantum black holes. In particular, the superposition of graphs (which is necessary to bring the analysis at the dynamical level) may be implemented by enriching the spin network
structure with data encoding the amount of link-entanglement between vertices, and using such data
to manipulate the combinatorial structure of the graph, analogously to what has been done for
random tensor networks\cite{Qi:2017ohu}. As far as an information-theoretic characterisation of black hole horizons is concerned, the illustrated techniques are for example expected to enable the derivation of a \virg{threshold condition} for the emergence of horizon-like surfaces in finite regions of quantum space, analogously to
the one obtained from the typicality approach to the study of the local behavior of spin networks\cite{Anza:2017dkd}. 

While the present article covered only a particular corner of the burgeoning field at the crossroads between quantum information and gravity, it is hoped our focused review might inspire further research and continue to motivate fruitful cross-fertilisation of methods and concepts between these two cutting-edge areas of theoretical physics, ultimately leading to their unification or confluence within a more fundamental theory yet to be discovered.


\begin{acknowledgments}
The authors would like to thank Goffredo Chirco, Daniele Oriti and Aron Wall for useful discussions and comments. E.C.~acknowledges funding from the DAAD, via the scholarship programme \virg{Research Grants - Short-Term Grants, 2021}, and thanks the Ludwig Maximilian University of Munich for the hospitality.
\end{acknowledgments}

\section*{Author declarations}
\subsection*{Conflict of interest}
The authors have no conflicts to disclose.

\section*{Data availability}
Data sharing is not applicable to this article as no new data were created or analysed in this study.

\section*{References}
\nocite{*}
\bibliography{aipsamp}

\end{document}